\documentclass[1p,number,sort&compress]{scrartcl}
\pdfpageattr {/Group << /S /Transparency /I true /CS /DeviceRGB>>}
\usepackage[utf8]{inputenc}

\usepackage{amsmath, amssymb, amsthm}
\usepackage{tikz}
\usetikzlibrary{arrows}
\usepackage{subfigure}
\usepackage{enumitem}
\usepackage{url}
\usepackage{here}
\usepackage{booktabs}

\newtheorem{theorem}{Theorem}

\newtheorem{defi}{Definition}

\newcommand{\N}{\mathbb{N}}

\newcommand{\R}{\mathbb{R}}

\renewcommand{\P}{{\ensuremath{\cal P}}}

\newcommand{\NP}{\ensuremath{\cal NP}}

\newcommand{\minProblem}[3][]{\begin{align}
\makebox[0pt][r]{#1\quad}\min\quad&#2&&\\\text{s.\,t.}%
\quad#3\end{align}}

\graphicspath {{figures/}}

\begin{document}

\title{Traffic signal optimization: combining static and dynamic models}

%\tnotetext[t1]{Extended version of results presented at the 15th meeting of the Euro Working Group on Transportation.}

\author{Ekkehard Köhler \phantom{andand} Martin Strehler \\[1ex] Brandenburg University of Technology, Mathematical Institute,\\
P.O.~box~10\,13\,44, 03013 Cottbus, Germany\\ \url{{ekkehard.koehler,martin.strehler}@b-tu.de}}
%\address{}

\date{\today}

\maketitle

\begin{abstract}
In this paper, we present a cyclically time-expanded network model for simultaneous optimization of traffic assignment and traffic signal parameters, in particular offsets, split times, and phase orders. Since travel times are of great importance for developing realistic solutions for traffic assignment and traffic
signal coordination in urban road networks, we perform an extensive analysis of the model. We show that a linear time-expanded model can reproduce realistic travel times especially for use with traffic signals and we verify this by simulation.  
Furthermore, we show how exact mathematical programming techniques can be used for optimizing the control of traffic signals. We provide computational results for real world instances and demonstrate the capabilities of the cyclically time-expanded by simulation results obtained with state-of-the-art traffic simulation tools.
\end{abstract}

%\begin{keyword}
%traffic signal optimization; traffic assignment; cyclically time-expanded network; flow over time; link performance; mixed integer programming
%\end{keyword}

\section{Motivation}\label{sec:intro}
% trc_signal_motiv.tex

\noindent Traffic signals can be seen as a backbone in the control of traffic flows in urban areas. Since the
appearance of the first signalized intersections over 100 years ago, the improvement of signal control
strategies has been an important subject for research. Friedrich~\cite{Friedrich02} and Papageorgiou et al.~\cite{papa}
provide overviews of the main research lines over the last years. 

Recently, we presented a model for
traffic signal coordination based on cyclically time-expanded networks (see~\cite{KS10,KS12,meinediss}). In
contrast to most of the previous mathematical approaches, we focused not only on optimizing the
offsets of signals, but also optimizing the implied traffic assignment. Each change in the signal parameters may influence
the travel times in the network. Consequently, road users will quickly adapt to a new signal coordination and
they will switch to faster routes if they suffer from the last intervention in the signal plans. The traffic assignment, i.e., the distribution of traffic in the road network, may change significantly.

Contrary to previous practical approaches based on non-linear models and heuristics, we focus on a model which allows for the use of exact programming techniques. Hence, proving optimality of a solution or bounding the gap between primal and dual solution was given a higher priority than modeling every effect of real traffic. Thus, this paper is particularly addressing the underlying theoretical questions of traffic signal optimization and the corresponding hardness.

In this paper, we present an extensive analysis of the cyclically time-expanded model. It is a hybrid between a static, i.e., time independent, and a dynamic, i.e., time dependent, network flow approach. Although this time-expanded model is based on a linear program, it can reproduce very realistic non-linear flow-dependent travel times for urban traffic networks with signal control. 
In particular, both typical convex link performance functions and the time-dependent behavior of traffic, necessary for traffic signal coordination, are realistically mapped in this model. Additionally, the propagation of platoons of cars that is essential for the minimization of waiting times in signal controlled road network is modeled. This is achieved by a lifting method. Whereas previous non-linear approaches use only a small number of variables to describe flow and travel times on a single road segment, we use a linear but high-dimensional model. Moreover, we do not use the common approach of linearizing an existing non-linear model. Instead, our linear model is built from scratch and it is compared to other approaches afterwards.

\paragraph*{Related Work} Traffic signal optimization is studied since the 1920s. Several models and methods have been proposed. Focussing only on the most theory-driven approaches, we refer to , e.g., Gartner et al.~\cite{gartner75}, Serafini and Ukovich~\cite{serafini} or Köhler et al.~\cite{bmbf-buch-ampeln}. These approaches rely on abstract models, but exact solution techniques can be applied. However, these approaches do not consider feedback to the traffic assignment. Other approaches use more sophisticated models, but afterwards heuristic algorithms have to be applied to carry out the optimization of signal settings. The most well-known example is the TRANSYT package, which is developed since in the 1960s beginning with the work of  Robertson~\cite{Robertson}. TRANSYT is based on an accurate traffic simulation and a genetic algorithm is used to optimize signal parameters.  

%Our approach is different to other mathematical models for signal optimization, e.g., those proposed by Gartner et al.~\cite{gartner75}, Serafini and Ukovich~\cite{serafini} or Köhler et al.~\cite{bmbf-buch-ampeln}, because  Our concept is also different to heuristic methods like TRANSYT (Robertson, 1969, see), since our model allows applying strict mathematical optimization techniques. 

Traffic assignment on its own is also a well studied problem. There are strict mathematical approaches using network flow theory. This branch of research started with the seminal work of Ford and Fulkerson~\cite{FF56} in 1956 and recent result especially focus on various aspects of traffic, e.g., a new concept of fairness in networks with congestion (Jahn, Möhring, Schulz, \& Stier Moses, see~\cite{Jahn05}) and flows over time with load dependent transit times (Köhler \& Skutella, see~\cite{KS05}). More application-oriented concepts include dynamic traffic assignment (DTA) techniques, see, e.g., Szeto \& Lo~\cite{szeto} and Chiu et al.~\cite{dtaprimer}, and simulation based solutions (Nagel \& Flötteröd, see~\cite{NagelFloetteroed2009IatbrResource}).

In contrast, the combined problem of traffic signal optimization and traffic assignment is rarely studied. Allsop and Charlesworth~\cite{allsop77} recognized the feedback between an optimized coordination and traffic assignment. They proposed an iterative approach where signal timings are optimized with TRANSYT. Afterwards, an equilibrium traffic assignment is computed. These steps are repeated until no change in the coordination occurs. Using heuristic methods iteratively, one may only hope to derive a local optimum. Even worse, this approach may lead to a decline of network performance as shown by Dickson~\cite{dickson}. Despite the strong interaction of signal coordination and traffic assignment, only some research was done in the area of integrated optimization of these two aspects of traffic. For example, Chiou~\cite{chiou05} presented a bilevel formulation based on the approach of Allsop and Charlesworth. Smith~\cite{smith06} also suggested a bilevel optimization, where the equilibrium property of the flow is preserved during the iterations. Recent results were also made by Bell and Ceylan~\cite{bell04} as well as Teklu et al.~\cite{teklu07} using genetic programming. Van den Berg et al.~\cite{Berg08} proposed a hybrid approach using both methods from optimal control and mixed integer linear programming. %In contrast, using traffic signals for actively influencing route choice is rarely studied. 
However, in a very recent paper, Smith~\cite{smith15} summarizes the situation as follows: 'At the moment, in practice, traffic signal timings are designed or optimised without systematically seeking to influence route choices beneficially.'

This short overview also reveals one of the main conflicts in traffic optimization. On the one hand, the practitioner favors very realistic and dynamic traffic models. However, these sophisticated models do usually not allow for strict mathematical optimization techniques for solving instances of relevant size. Instead, heuristics, e.g., genetic algorithms or line search strategies in non-convex settings, have to be used. Yet, they are very sensitive to local optima and neither provide guarantees on the gap towards the optimal solution nor on the convergence ratio. Thus, one may only hope to improve the present solution, but the actual optimal solution remains unknown. On the other hand, strict mathematical optimization approaches like linear programming often require several assumptions on the problem formulations which lead to very simplified traffic models. However, these strict mathematical strategies often provide much more insight in the underlying problems, e.g., via the dual problem formulation. That is, out of the solution we can directly derive quantified suggestions for an improvement of the network performance. Therefore, one has to find a compromise between a qualified optimality result and the model's accuracy.

\paragraph*{Our contribution} In this paper we aim at the crossover between static and dynamic models for traffic assignment. Our main result is a combined optimization approach for simultaneous traffic assignment and traffic signal coordination with realistic load dependent travel times which still allows applying strict mathematical optimization techniques.

Firstly, we shortly present static and dynamic traffic models and introduce the main concept of our cyclically time-expanded network model. In Section~\ref{sec:performance}, we show that this new discretized and linearized model yields load dependent travel times which resemble main properties of common link performance functions. Moreover, our model is also capable of capturing platoons of cars of quickly varying density as well as changing phases of traffic signals. We verify the applicability of our model by simulation with state-of-the-art simulation tools, namely VISSIM and MATSim.
%Nevertheless, our model is still linear. Thus, traffic assignments 
%can be computed very efficiently. 
Furthermore, we present an extension of our cyclically time-expanded model in Section~\ref{sec:opt}, which allows for the simultaneous optimization of traffic signal parameters to design fixed time or time of day signal timings. Whereas the previous version of our model was only capable of optimizing offsets, we extend our approach to the simultaneous optimization of split times and phase orders. Afterwards, we study solutions of real-world instances in Section~\ref{sec:results}. Finally, we use the traffic simulation MatSim to investigate the price of anarchy for our scenarios, i.e., the gap between system optimum and user equilibrium.

\section{A Cyclically Time-Expanded Model for Traffic Assignment}\label{sec:model}

% trc_signal_model.tex

\subsection{Static and dynamic models for traffic assignment}

\noindent Intersections and connecting roads can be represented by the set $V$ of nodes and the set $A$ of links of a (directed) network $G=(V,A,u)$, where $u:A\to \R_0^+$ assigns a capacity to each of the arcs in the network. In a static network flow model, we assign flow values $f:A \to \R_0^+$, such that capacities are not exceeded and flow conservation is satisfied. That is, the amout of incoming flow equals the amount of outgoing flow. First results on efficiently computing maximum flows were obtained by Ford and Fulkerson~\cite{FF56} already in the 1950s. Due to the abundance of applications ranging from traffic flows to image processing, numerous algorithms and modifications have been developed.

However, many applications require that the flow is a function of time, i.e., flow values are changing over time. Generally, a (traffic) \emph{flow over time} or {dynamic flow} can formally be defined as a function $f:A\times\left[ 0,T\right) \to \R_0^+$ on a (directed) network $G=(V,A,u)$, where  $\left[ 0,T\right)$ is the time interval under consideration. Flow particles entering $e\in A$ at time $t$ arrive at the head of $e$ at time $t+t_e(f(e,t))$, where $t_e:\R_0^+\to \R_0^+$ is the travel time function or \emph{link performance function} of link $e$. The amount of flow that passes a link $e$ can be calculated by $\int_0^T f(e,t) dt$. Furthermore, \emph{flow conservation} has to ensure that flow cannot leave a node before it arrives there. Additionally, one may assume that flow can be stored in a node for a certain time span representing queues at intersections. In detail, for load-indepentent travel times $t_e$, for each $\tau \in \left[ 0,T\right)$, and each non-terminal $v\in V$, $\sum_{e\in\delta^-(v)} \int_0^{\tau-t_e} f(e,t) dt \ge \sum_{e\in\delta^+(v)} \int_0^{\tau} f(e,t) dt$, where $\delta^-(v)$ and $\delta^+(v)$ refers to the set of incoming and outgoing arcs of $v$.

Flows over time are widely studied in the mathematical literature; Skutella provides a good overview of recent results~\cite{Skutella2009}. Many dynamic flow problems with travel times independent of the flow can be solved using \emph{time-expanded networks}. Already Ford and Fulkerson introduced time-expanded networks in their seminal work on network flow theory~\cite{FF56}. To create an expanded network out of a simple network, for every node, several copies of this node are added to the graph, one for each desired time step. These nodes are connected by arcs, where the various copies of the vertices are connected according to the travel times of the original arcs. Additionally, arcs connecting consecutive copies of the same node model waiting at this node. Yet, time-expanded networks are rather inefficient if the time horizon $T$ is large, since the number of time steps determines the number of network copies that have to be provided.

Ford and Fulkerson's time-expanded networks were also applied in traffic flow theory, often called \emph{space-time expanded networks} (STEN) in this context~\cite{YangMeng1998}. \emph{Dynamic traffic assignment} (DTA) combines flow over time with variable departure times and non-linear link performance functions. It is used to model flow dependent travel times and to compute user equilibria according to Wardrop’s principle. This approach allows the usage of solving techniques of Control Theory or Calculus of Variations. However, even for medium size networks it is very difficult to derive analytical solutions. Therefore, to be able to compute at least numerical solutions in an iterative approach, time is discretized again and several other parameters like the number of routes are limited.

\subsection{The cyclically time-expanded network}\label{sec:cycmodel}

\noindent For traffic signals and their coordination, a time-dependent model, capable of describing the time-offset between consecutive intersections, is a vital ingredient. The impact of \emph{red} signals on the travel times is essential. 

Link performance functions, as suggested by the Bureau of Public Roads (BPR), are a widely accepted approach to model load-dependent travel times. However, these static link performance functions describe traffic in a rather statistical manner, i.e., they  support an a priori estimation of the expected average delay. Especially without standard deviations, it is hardly possible to deduce the travel time of an individual driver. Hence, link performance functions do not provide enough information to be used directly for traffic signal coordination.  Furthermore, to the best of our knowledge, approaches like DTA are not used to model dynamic flow in realistic large scale scenarios on such a fine timely level as needed for traffic signals.

Indeed, one should look into the causes for flow dependence of travel times in urban areas. In inner-city road networks, the individual behavior of road users is of minor importance for the travel times. In comparison, traffic signals and the rather low speed limit have a much greater influence on travel times. Thus, for urban
networks, we suggest to decompose the travel time into \emph{pure transit time} and \emph{waiting time} at the
intersections. In the following, we assume that the \emph{transit time is constant} and the \emph{flow-dependent
component of travel time is caused only by the delay} at traffic signals.

With constant transit times, time-expanded networks come into play again. Even better, since pretimed
traffic signals have a periodic behavior, it is not necessary to use a full time horizon expansion.
Instead, we suggest a \emph{cyclic time-expansion} where we expand only a time interval of size of the cycle
time $\Gamma$ of the traffic signals and use only $k\in \N$ time steps of size $t=\frac{\Gamma}{k}$.

\begin{defi}[Cyclically time-expanded network]
 Let $G=(V,A,u)$ be a network with capacities $u:A\to \N$ and non-negative integral transit times $t_e$ for each $e\in A$. For a given number $k$ of time steps of length $t=\frac{\Gamma}{k}$, the corresponding \emph{cyclically time-expanded network} $G^T=(V^T,A^T,u^T)$ is constructed as follows.
 \begin{itemize}
  \item For each node $v\in V$, we create $k$ copies $v_0,v_1,\dots,v_{k-1}$, thus $V^T=%\left
  \{ v_t| v \in V,t\in\{0,\dots,k-1\}
  %\right
  \}$.
  \item For each link $e=(v,w)\in A$, we create $k$ copies $e_0,e_1,\dots,e_{k-1}$ where arc $e_t$ connects node $v_t$ to node $w_{(t+t_e) \mod k}$. These arcs are called \emph{transit arcs} and $e_t$ has capacity $u(e_t):=u(e)$.
  \item Additionally, there are \emph{waiting arcs} from $v_t$ to $v_{t+1}$ $\forall v\in V$ and $\forall t \in \{0,\dots,k-2\}$ and from $v_{k-1}$
to $v_0$.
 \end{itemize}
\end{defi}

Please note that there is almost no difference between transit arcs and waiting arcs in the model. On
waiting arcs one only moves in time whereas on transit arcs one moves in space and time. For all arcs
the cost is the travel time on this arc. Thus, waiting time is travel time on waiting arcs. Hence, both kinds
of arcs are treated in the same way and we do not need to explicitly distinguish between them in the
modeling of traffic assignment. In practice, we will compute the transit time via the speed limit and the length of the road segment, whereas the waiting time on a waiting arc is simply determined as the duration of a time step in the expansion. 

\begin{figure}[ht]
 \centering
\begin{picture}(0,0)%
\includegraphics{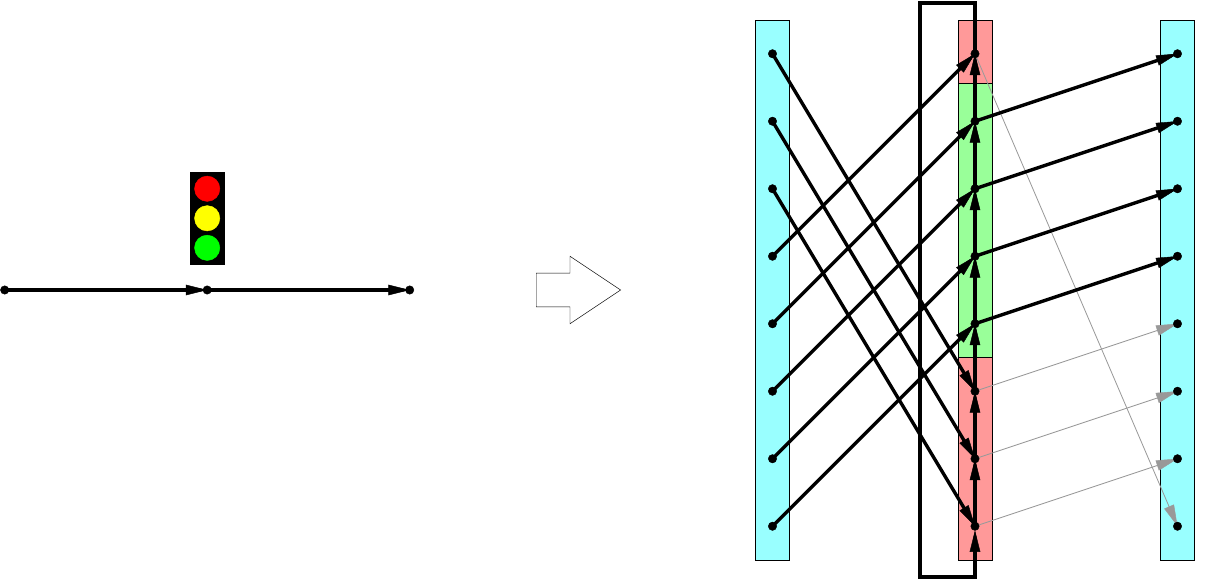}%
\end{picture}%
\setlength{\unitlength}{1066sp}%
\begingroup\makeatletter\ifx\SetFigFont\undefined%
\gdef\SetFigFont#1#2#3#4#5{%
  \reset@font\fontsize{#1}{#2pt}%
  \fontfamily{#3}\fontseries{#4}\fontshape{#5}%
  \selectfont}%
\fi\endgroup%
\begin{picture}(21848,10308)(-12832,-9415)
\put(-11399,-4861){\makebox(0,0)[lb]{\smash{{\SetFigFont{9}{10.8}{\rmdefault}{\mddefault}{\updefault}{\color[rgb]{0,0,0}$t_{e_1}=3$}%
}}}}
\put(-7874,-4861){\makebox(0,0)[lb]{\smash{{\SetFigFont{9}{10.8}{\rmdefault}{\mddefault}{\updefault}{\color[rgb]{0,0,0}$t_{e_2}=1$}%
}}}}
\put(9001,-8761){\makebox(0,0)[lb]{\smash{{\SetFigFont{9}{10.8}{\rmdefault}{\mddefault}{\updefault}{\color[rgb]{0,0,0}$t=0$}%
}}}}
\put(9001,-7561){\makebox(0,0)[lb]{\smash{{\SetFigFont{9}{10.8}{\rmdefault}{\mddefault}{\updefault}{\color[rgb]{0,0,0}$t=1$}%
}}}}
\put(9001,-6361){\makebox(0,0)[lb]{\smash{{\SetFigFont{9}{10.8}{\rmdefault}{\mddefault}{\updefault}{\color[rgb]{0,0,0}$t=2$}%
}}}}
\put(9001,-5161){\makebox(0,0)[lb]{\smash{{\SetFigFont{9}{10.8}{\rmdefault}{\mddefault}{\updefault}{\color[rgb]{0,0,0}$t=3$}%
}}}}
\put(9001,-3961){\makebox(0,0)[lb]{\smash{{\SetFigFont{9}{10.8}{\rmdefault}{\mddefault}{\updefault}{\color[rgb]{0,0,0}$t=4$}%
}}}}
\put(9001,-2761){\makebox(0,0)[lb]{\smash{{\SetFigFont{9}{10.8}{\rmdefault}{\mddefault}{\updefault}{\color[rgb]{0,0,0}$t=5$}%
}}}}
\put(9001,-1561){\makebox(0,0)[lb]{\smash{{\SetFigFont{9}{10.8}{\rmdefault}{\mddefault}{\updefault}{\color[rgb]{0,0,0}$t=6$}%
}}}}
\put(9001,-361){\makebox(0,0)[lb]{\smash{{\SetFigFont{9}{10.8}{\rmdefault}{\mddefault}{\updefault}{\color[rgb]{0,0,0}$t=7$}%
}}}}
\end{picture}%
 
 \caption{A small network with two links and one traffic signal (left) and its cyclic time-expansion for $k=8$ (right). The thin, gray arcs starting at the red phase
of the signal are switched off and, thus, they have zero capacity.}\label{fig:example}
\end{figure}
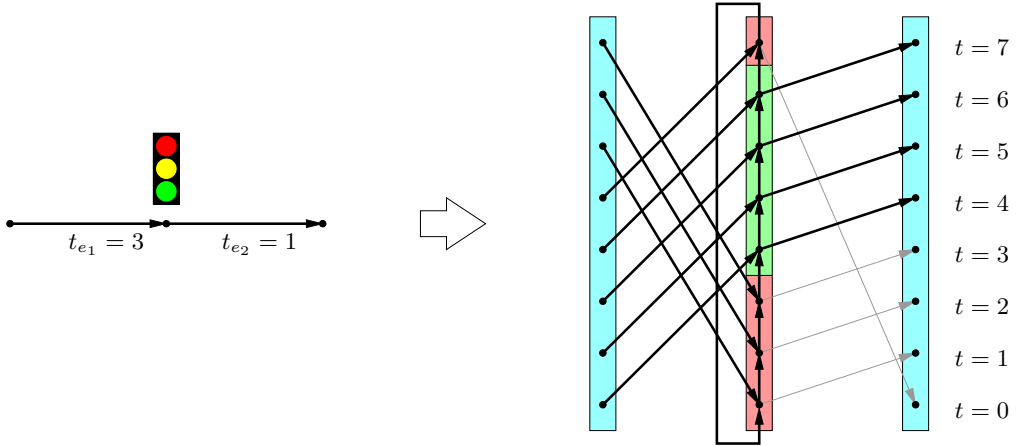

Now, traffic signals with their \emph{green} and \emph{red} phases can be modeled by setting the capacity of a transit arc
starting during the \emph{red} phase to zero, i.e.,~$u(e_t):=0$ for some $t$. A simple example of a cyclically time-expanded
network consisting of two links and one traffic signal is shown in Figure~\ref{fig:example}. Additionally,
capacities of the waiting arcs correspond to the maximum queue length on a link and they can be chosen
accordingly. This also allows the modeling of vehicle spillback. If the capacity of a waiting arc is exhausted and the signal is still \emph{red}, no more flow can enter the corresponding transit arc due to the flow conservation constraint. Thus, flow on the upstream link has to use the waiting arc there, although the associated traffic light might be \emph{green}.

Consequently, the cyclically time-expanded network model is a hybrid inbetween static and dynamic models. On the one hand, it provides several advantages of static flow models. Due to its linearity and limited size, standard approaches for computing (static) multi-commodity flows can be applied. On the other hand, it provides sufficient time resolution to account for traffic signals. Yet, only a narrow time slice is considered, hence it is not a fully dynamic model.

\section{Inherent Link Performance in the Cyclically Time-Expanded Network}\label{sec:performance}

% trc_signal_performance.tex
\noindent
So far, we have introduced a model with constant travel times on transit arcs and waiting arcs. Consequently, travel times are independent of the
flow on the link. From a practitioner's point of view, this might seem too restrictive in the first
moment. However, there is in fact a flow-dependent behavior and it is hidden in the time expansion. We will
now study this flow dependency by an analysis of our model for a single road with a traffic signal.
Furthermore, we support the observations by a simulation of traffic flow on this link. As a main insight, we will see that our high-dimensional linear model can reproduce the same effects as a classical low-dimensional non-linear traffic flow model.

In the cyclically time-expanded model, we can look at traffic flow in two ways. \emph{Firstly}, we can
examine each of the copies of an original link and the adjacent waiting arcs on their own. This provides a
dynamic and detailed view on the traffic flow and we can observe flow particles moving through the
network in a timely fashion. Considering only one road segment, which is modeled by $k$ transit arcs and $k$ waiting arcs, we have a $2k$-dimensional linear relation between flow and travel time.

\emph{Secondly}, we can virtually contract this flow over time back to the original
link, i.e., we regard all copies of the links and the corresponding waiting arcs as a single link and compute
the average travel time over the whole time horizon. For a single link, we can imagine this as a projection of the $2k$-dimensional space, which yields a piecewise linear relation between total flow and average travel time. In that way, we have a more static view on traffic
flow on this contracted link and temporal details are lost, but we can compare the travel time to other
static models, now.

In the following, we will use the first point of view to compute the average travel times for the second
perspective. We will see that travel time on a contracted link depends on numerous parameters, e.g., the
distribution of incoming flow values over time. Furthermore, flow particles on the same (contracted) link
at different times will experience different travel times which is very obvious in the time-expanded
model. We use \emph{link performance} to denote the average travel time on a contracted link. However, we will not
present a closed function for computing travel times due to the enormous number of parameters. Nevertheless, we use the term \emph{inherent link
performance function} according to related static traffic assignment models with flow dependent travel
times. In the following, we fix most of the parameters and restrict the inherent link performance function
to a small subset of its domain to at least demonstrate the most important properties of this implicit link
performance in the cyclically time-expanded model.

\subsection{Evolving flow-dependent waiting times}

\noindent
We consider a very small network which consists of a single road with a traffic signal in the middle
similar to the network in Figure~\ref{fig:example}. Assume that a set of flow values is assigned to the incoming transit arcs in the cyclically time-expanded
network of this scenario and a fixed signal setting is given. The flow values of the other arcs are chosen such that flow conservation is fulfilled and flow is assigned to a waiting arc if and only if the capacity of the outgoing transit arc at this node is exhausted. That is, flow units only wait when the outgoing road is already blocked by other flow units.  

\begin{figure}[ht]
 \centering
 \includegraphics[width=6cm]{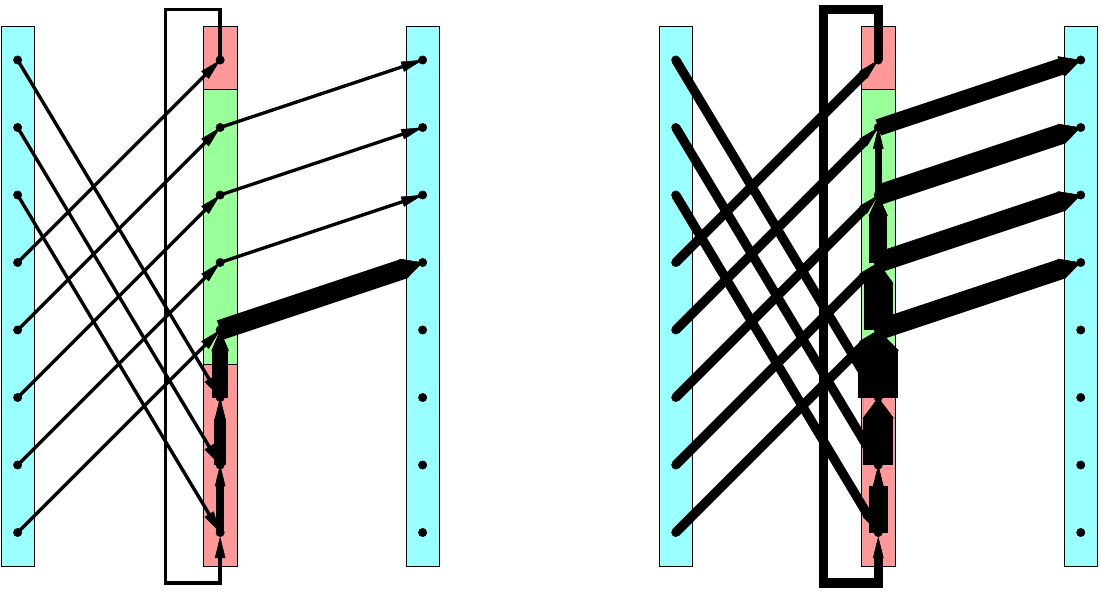}
 \caption{The same single road network with two different flows. The thickness of the arcs depicts the flow value. On the left side, incoming flow is low. All
flow accumulated during the \emph{red} phase can leave the signal directly when it becomes \emph{green}. On the right side, the incoming flow is tripled.
Thus, much more flow is accumulated on the waiting arcs. Furthermore, due to the capacity bound of the outgoing street, only a fraction of flow can leave on the first outgoing transit arc and more waiting arcs have to be used. This yields an increase of waiting time by a factor of 4.8.}\label{fig:flow}
\end{figure}

The total waiting time in this small network can be determined by summing up all flow value of the waiting arcs and multiplying the result by the length of one time step. 
Now, we increase the flow values on the incoming transit arcs. This will increase the waiting time, because more flow will be assigned to the waiting arcs.
Furthermore, due to the bounded capacities, the flow units on the waiting arcs may not leave completely
at the first green outgoing transit arc if this link does not provide enough capacity. Instead, the flow will have to
use more waiting arcs until the accumulated flow is drained off. This relation is illustrated in Figure~\ref{fig:flow}.
Therefore, if the incoming flow is raised linearly on all copies of the incoming transit arc, then the growth of the waiting
time will not be linear but quadratic. More precisely, the obtained function is piecewise linear, but
converges to a quadratic function if the length of the time steps tends to zero.

\begin{figure}[ht]
 \centering
 \includegraphics[width=5cm]{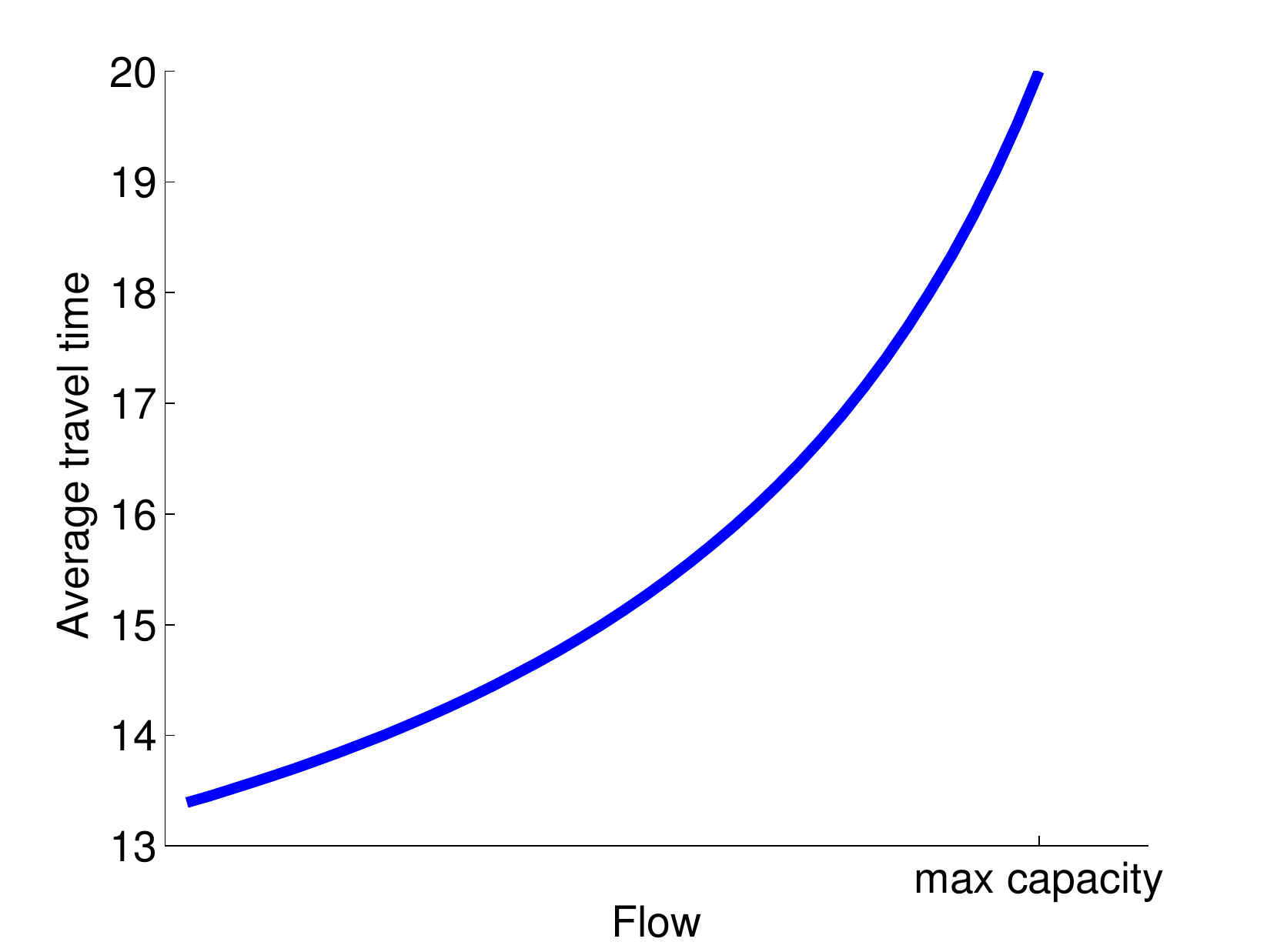}
 \caption{Relation between average travel time and flow value on a single link in the cyclically time-expanded network. Incoming
traffic is uniformly distributed over time.}\label{fig:traveltimes}
\end{figure}

In Figure~\ref{fig:traveltimes}, we present the relation between flow and average waiting time in the cyclically time-expanded network with a finer time granularity. A cycle time of 60 seconds and 60 time steps are used. The incoming traffic is uniformly distributed on all transit arcs, a traffic signal which is \emph{red} for 20 seconds is put at the end of the road, and a free speed travel time
of 10 seconds is assumed. Additionally, a capacity reduction from two lanes to one lane at the traffic
signal was used. We now compute the average travel time in this scenario with respect to the flow value.
The resulting travel time in Figure~\ref{fig:traveltimes} demonstrates the capability of our waiting arc model: although
using only constant travel times the inherent, implicit link performance functions of the model are not
linear. Even better, the implicit link performance resembles a common standard link performance function. Please note that Figure~\ref{fig:traveltimes}
visualizes the average travel time of a flow particle on the ‘contracted’ link. The individual travel time of
a single flow unit depends on its arrival time at the traffic signal. It ranges from 10 seconds for flow units
arriving at \emph{green} with no waiting queue at the signal up to 30 seconds for road users arriving at the
beginning of the \emph{red} phase.

This example implies that our model uses flow dependent travel times even tough they are not explicitly specified. But within this analysis, we assumed traffic to be uniformly distributed over time. Obviously, this is an unrealizable assumption when the steady flow is disturbed by traffic signals. Hence, we extend our study to flows with quickly changing traffic density.

\subsection{Travel times for platoons of cars}

\noindent
In inner-city traffic, one can often observe platoons of cars. Traffic signals cause an accumulation of cars. During one cycle of the signals, there are periods with cars nose-to-tail and periods with empty streets. In the cyclically time-expanded network, platoons of cars can be easily modeled by different flow values on
the copies of each particular transit arc. Some copies may be used at full capacity, other arcs may not be used
at all. Varying flow values can be interpreted as platoons of different lengths and densities. Thus, platoons are also created, splitted, merged, compressed, or stretched in our model in a very natural and self-acting manner.

Even in our small example of a single road with one traffic signal, a platoon can change the
characteristics of the average waiting time dramatically. Let us describe a platoon of cars by its length
and its density. Suppose for the moment the platoon to be homogeneous, i.e., the density is not varying inside the platoon. Therefore, the platoon length can be
considered to be proportional to the average link flow. For simplicity, we measure the length of a platoon as time elapsing between passing of head and tail of this platoon. This can be easily transformed to length in meter or number of cars, when speed or distance between cars is known.  Average travel time now depends on two parameters: the length of the platoon and the arrival time of the head of the platoon at the traffic signal.

\begin{figure}[ht]
 \centering
 \input{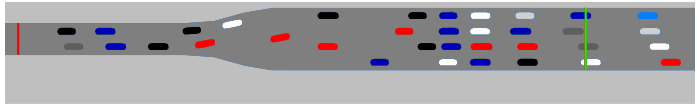}
 \caption{The described scenario was modeled in VISSIM (screenshot detail). Travel time is measured between $A$ shortly after the upstream signal and $B$  shortly after the actual signal. Due to the first signal, platoons are generated.}\label{fig:scenario}
\end{figure}

To emphasize the occurring effects even more, we
change the \emph{green} period to 30 seconds and the capacity of the outgoing road is set to twice the capacity of
the incoming road, e.g.,~the number of lanes are doubled. The scenario is visualized in Figure~\ref{fig:scenario}.

\begin{figure}[ht]
 \centering
 \includegraphics[width=5cm]{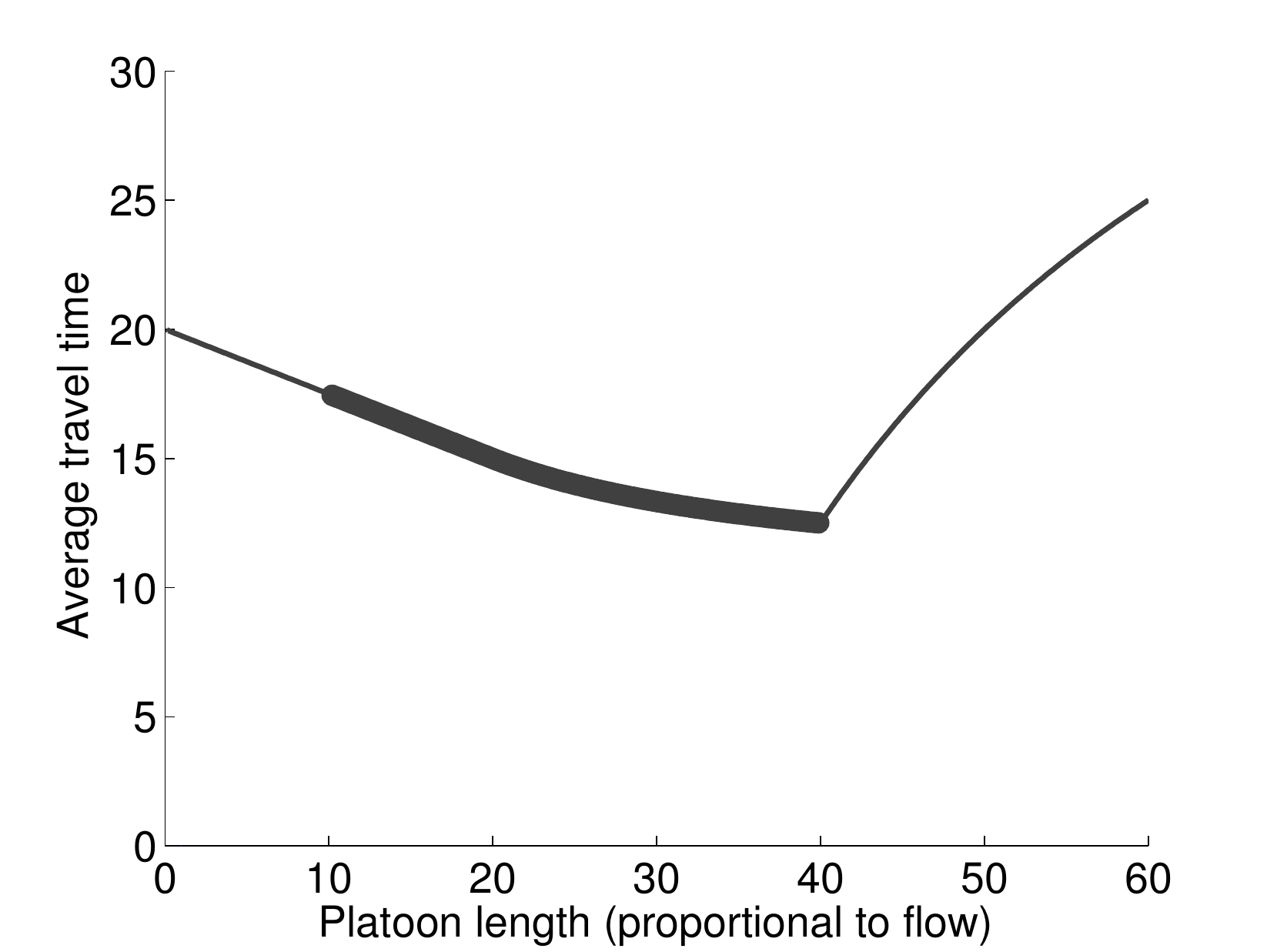}\hspace{2cm}\includegraphics[width=5cm]{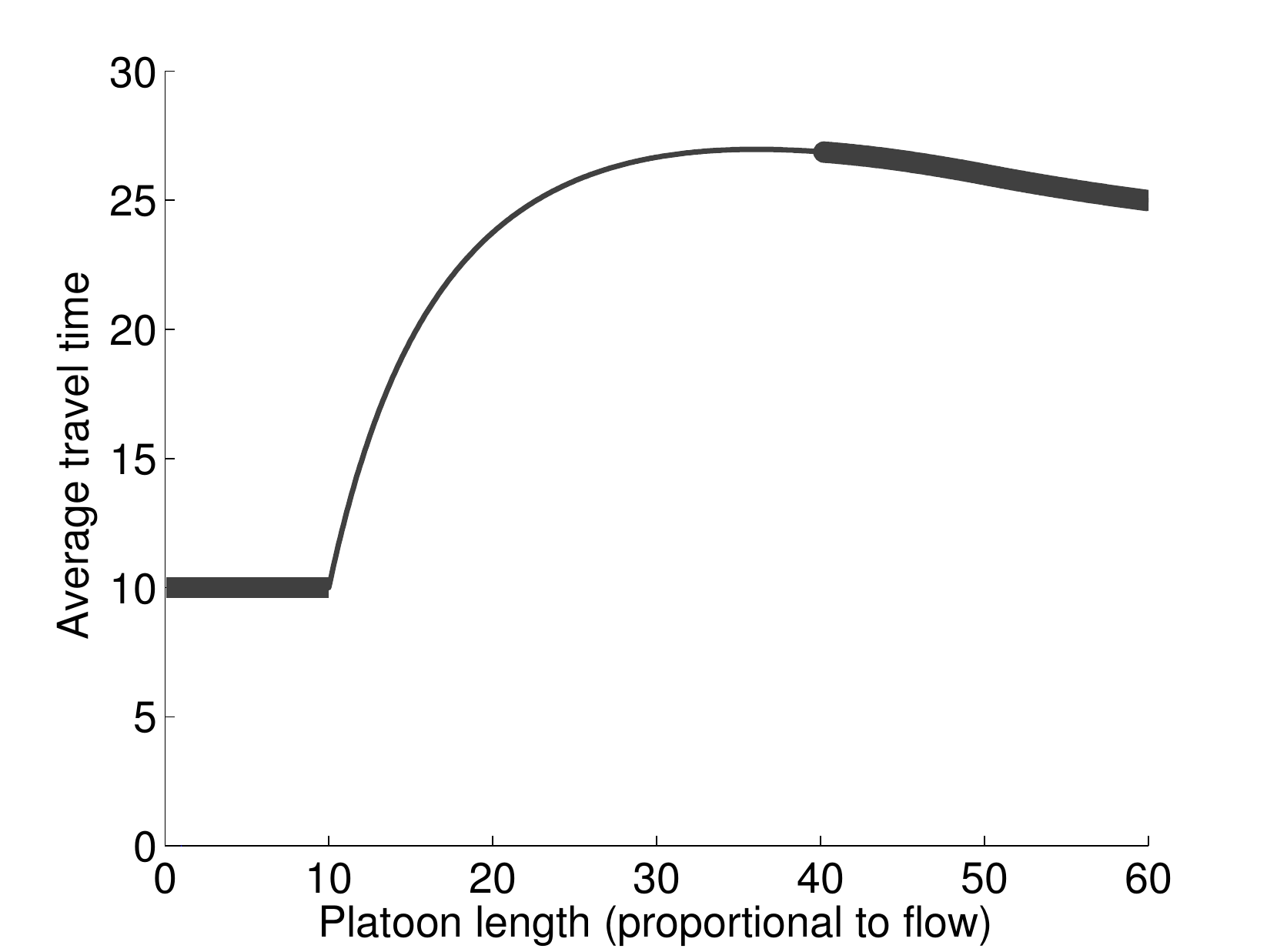}
 \caption{The plots show the inherent link performance function for a platoon of cars in the cyclically time-expanded
model for two different offsets. The thick line also visualizes the \emph{green} period, i.e., that part of the platoon arriving at a \emph{green} signal. In the left diagram, the head of the platoon is arriving 10 seconds before the signal turns \emph{green}. Due to the higher capacity of the outgoing
road, that is, the cars may leave on parallel lanes, the platoon is densified. The first car of the platoon has to
wait for the longest time, subsequent cars may even pass the signal without stopping. On the right side, the first flow particle arrives 20 seconds after the
signal turned \emph{green}. Thus, small platoons can pass without stopping, long platoons are splitted by the traffic signal which turns \emph{red}
10 seconds after the head of the platoon has passed.}\label{fig:traveltimes_platoon}
\end{figure}

In Figure~\ref{fig:traveltimes_platoon}, the calculated average travel time is plotted versus the platoon length. In Figure~\ref{fig:traveltimes_platoon}~(left),
the first flow unit arrives at the intersection 10 seconds before the signal turns \emph{green}. In Figure~\ref{fig:traveltimes_platoon}~(right), the head of the platoon arrives 20 seconds after the signal turned \emph{green}. As a consequence, the traffic signal is used to densify the traffic in the first setting. In the second setting, long platoons are split into two smaller platoons, of which the second one has to wait at the signal. The different characteristics of the functions in Figure~\ref{fig:traveltimes_platoon} demonstrate the impact of these two offsets on our travel times. %\comment{bildverlabelung!}

The implicit travel times for platoons show interesting or even unexpected properties. The travel times are decreasing or concave in some intervals, so they do not resemble standard static link
performance functions found in traffic literature 
%(e.g. Sheffi~\cite{Sheffi}, Gartner et al.~\cite{GartnerMesserRathi}, and Ortúzar \&
%Willumsen~\cite{Ortuzar}).
(e.g.,~\cite{Ortuzar,GartnerMesserRathi,Sheffi}).
There, link performance functions are usually assumed to be convex and monotonically
increasing with the amount of flow on the link. However, this is no inaccuracy or disadvantage of our model. As we will show,
these characteristics of travel time functions can in fact be observed in state-of-the-art traffic
simulation tools which supports our model.

\begin{figure}[ht]
 \centering
 \includegraphics[width=5cm]{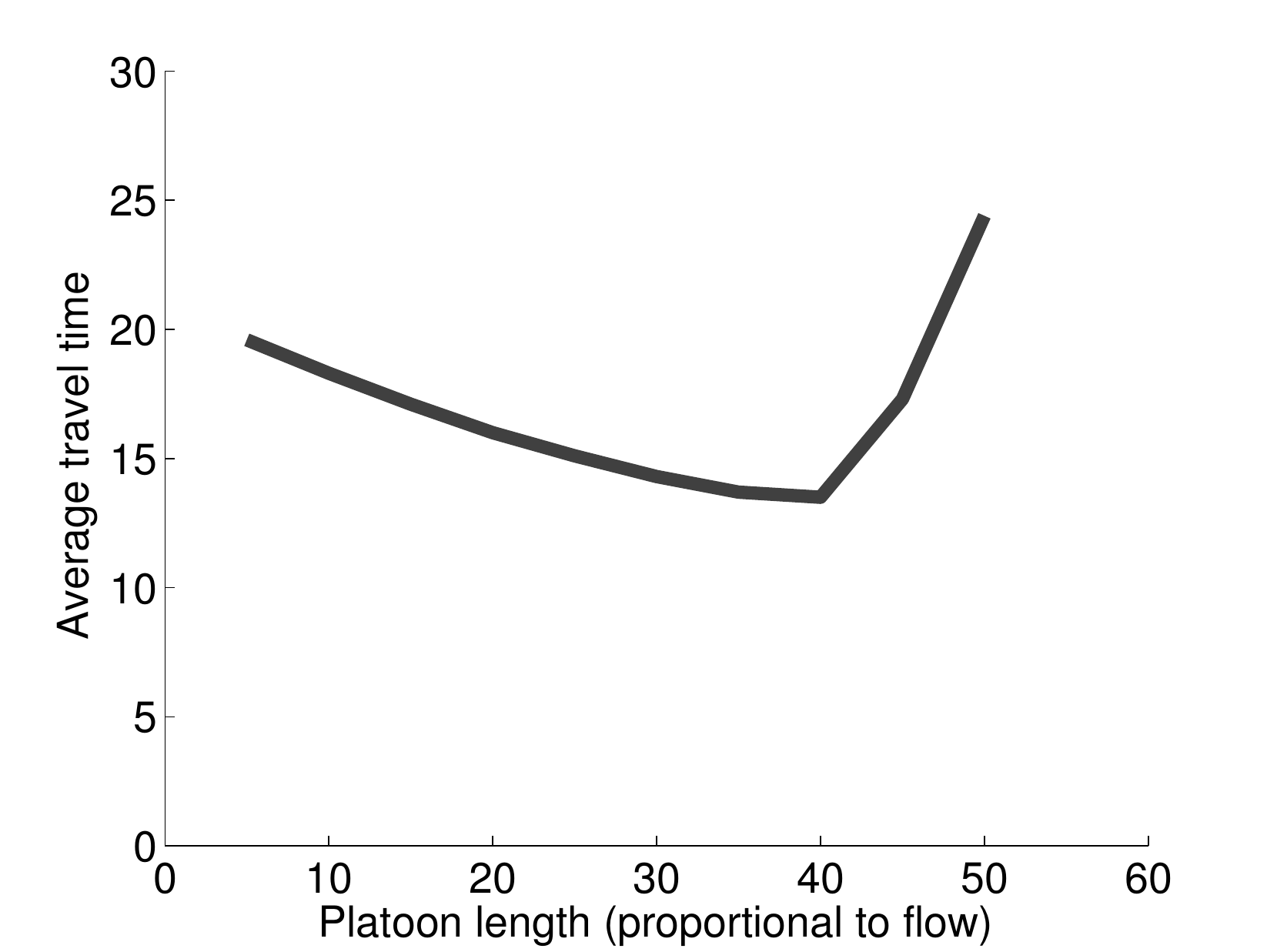}\hspace{2cm}\includegraphics[width=5cm]{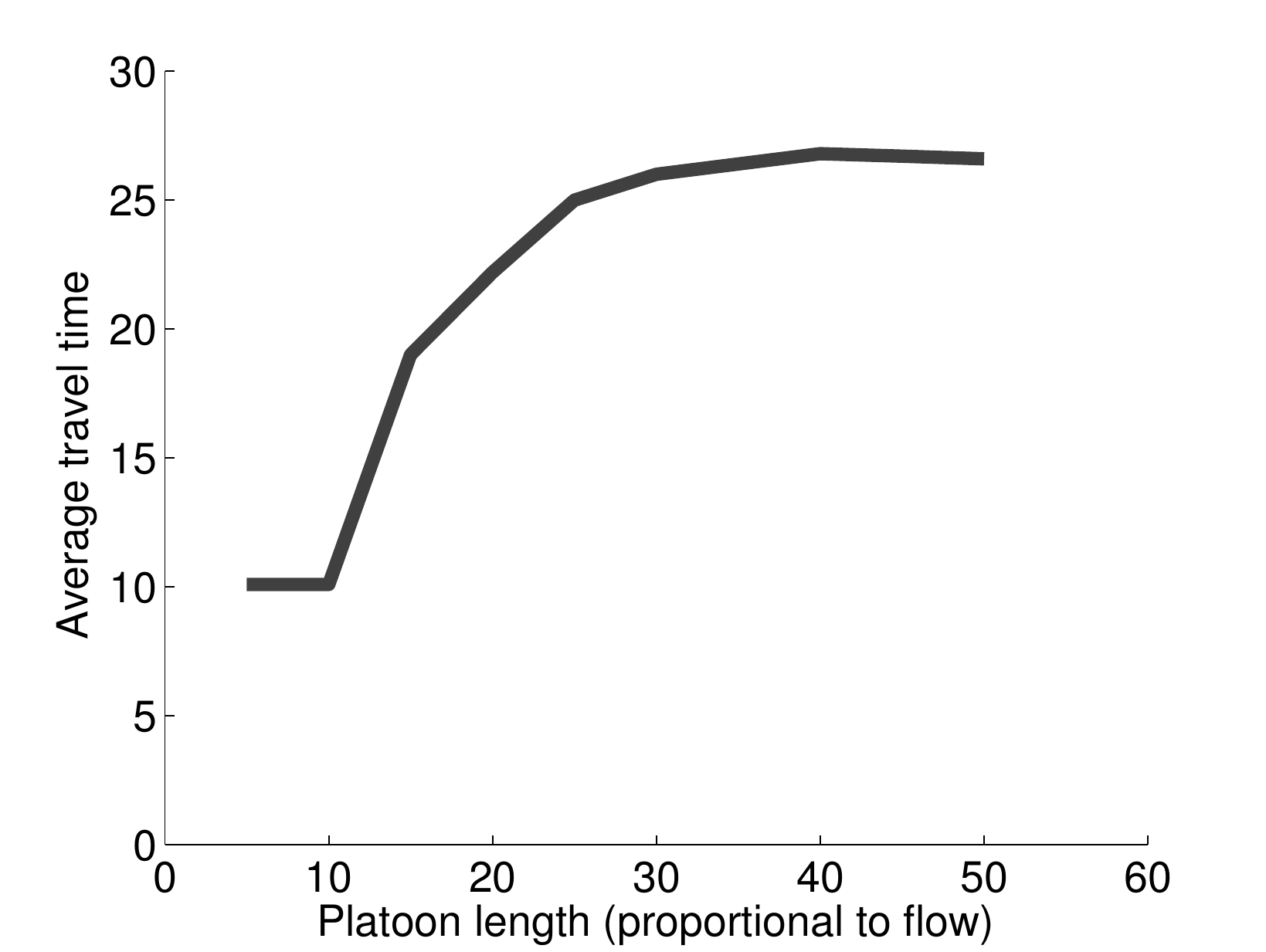}
 \caption{Link performance for the same scenario as in Figure~\ref{fig:traveltimes_platoon}, but now simulated and measured with VISSIM. Note that only the
average travel time is displayed. The actual time depends on the position of the car within the platoon.}\label{fig:traveltimes_sim}
\end{figure}

To evaluate the travel times computed by the cyclically time-expanded model, we use two well
established simulation tools, namely VISSIM and MATSim. VISSIM, developed by ptv group, is a state-of-the-art micro-simulation, featuring nearly every aspect of urban traffic~\cite{vissim}. MATSim, mainly developed by
TU Berlin and ETH Zurich, is an agent-based simulation capable of computing large scale traffic simulations. MATSim uses a simpler queuing model, but it
supports the computation of equilibrium assignments for thousands of traffic participants~\cite{webseite-matsim}. Figures~\ref{fig:scenario}  and~\ref{fig:traveltimes_sim}
present the described scenario and the corresponding measured travel times, respectively. To create the platoons, we use an additional upstream traffic signal (in front of $A$ in Figure~\ref{fig:scenario}). The arrival time of the platoon at the signal of interest can be influenced by changing the offset between both signals. The length of the platoon is determined by the \emph{green} period of the upstream signal.

As one can clearly see, the
predicted travel times in Figure~\ref{fig:traveltimes_platoon} fit remarkably well to the simulated travel times  in Figure~\ref{fig:traveltimes_sim} even without a careful
calibration of our model. Hence, one can conclude that the cyclically time-expanded model can also capture time dependency of flow on a very fine level, especially for platoons of cars and traffic signals. 
In contrast, static link performance functions are the best choice for describing traffic flow in rural areas or on highways, but they are not accurate enough for signalized inner-city traffic. This also affects user equilibrium flows. Here, the common assumptions on convexity and monotonicity are crucial for the existence of unique equilibria. 

\begin{figure}[ht]
 \centering
 \includegraphics[width=8cm]{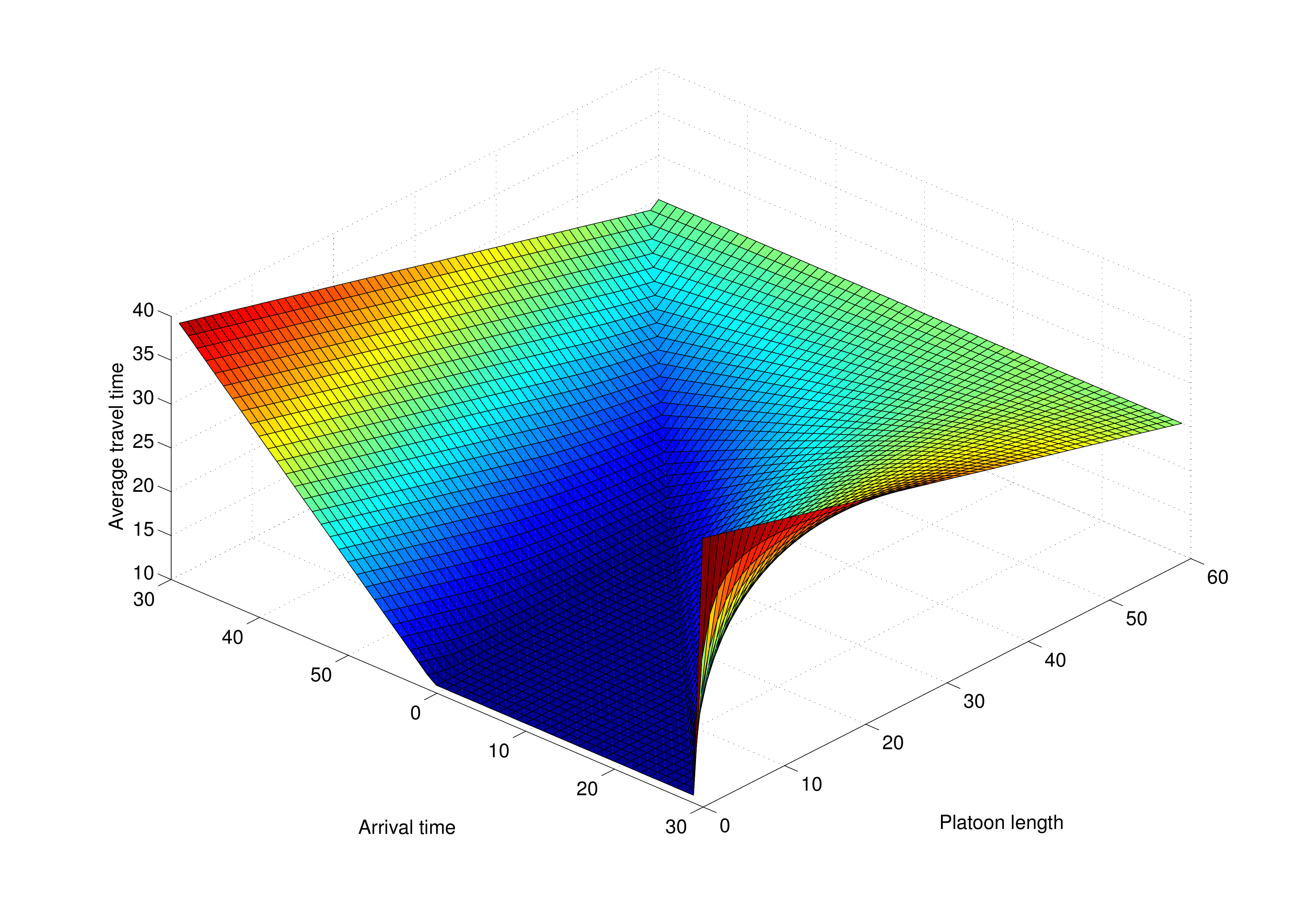}
 \caption{Calculated average travel time for a road user in a platoon with respect to platoon length and arrival time at the traffic signal.
The signal turns \emph{green} at $t=0$ and \emph{red} at $t=30$. The inherent link performance functions of Figure~\ref{fig:traveltimes_platoon} are obtained as profile for $t=50$ and $t=20$, respectively. Note that for fixed platoon length one obtains waiting time functions very similar to those used by Wünsch~\cite{Wuensch}.}\label{fig:traveltimes_complete}
\end{figure}

However, we cannot present a closed formulation of travel times for platoons and signals here, since this depends on various parameters, e.g., number, length and
density of platoons, traffic signal settings, and capacity changes of the link. If most parameters are fixed, the
average travel time of a flow unit of a single platoon in our scenario is depicted in Figure~\ref{fig:traveltimes_complete} with respect to
both parameters platoon length (i.e., traffic flow) and arrival time at the intersection (i.e., offset of the
signal).

Summarizing, our cyclically time-expanded network provides a piecewise linear approximation of classical non-linear link performance functions, but it can also keep track of platoons of cars and varying traffic density. To avoid any misunderstanding, please note that the above results are just used to compare our model to other non-linear approaches. The non-linear link performance functions that were studied in this section are not used in the remainder of this paper. All further results are obtained by using the original cyclically time-expanded network with its constant transit times.

\section{Traffic Signal Optimization in the Cyclically Time-Expanded Model}\label{sec:opt}

% trc_signal_opt.tex
\noindent
Up to now, we analyzed properties of the cyclically time-expanded model. In the following,
we show how our model can be used for traffic signal optimization. While we only considered offset optimization in our previous paper~\cite{KS10}, we now present an extended approach which allows simultaneous optimization of traffic assignment, traffic signal offsets, split times and phase order.

%\comment{However, changes in assignment lead to different demands at single traffic lights. Thus, adjusting split times could improve the capacity of an intersection. Optimizing the phase order of large junctions with many lanes may also improve the coordination significantly. }

\subsection{Efficient computation of traffic assignments}

\noindent
Building up on the cyclically time-expanded network, we can now
consider the traffic assignment problem within this framework.
Although flow can be considered to travel in a time-dependent manner
through the cyclically time-expanded network, one should rather see
this model as a static model that just captures some time-dependent
aspects of traffic flow.  Due to the cyclic repetition of the
vertex and arc copies, a flow particle traveling through this network
can be seen as a representative of a whole set of temporally repeated
particles at every multiple of the cycle time.

There is a closely related interpretation of static
network flow for traffic networks.  In this interpretation one
considers a flow-carrying path in the static network as a mapping of a
corresponding amount of flow particles traveling over time through the
traffic network at the corresponding flow rate.  In other words, a
flow-carrying path in the static network represents a constant rate of
flow on this path in the `real' time-dependent traffic network.

This interpretation suggests how to put together the two models, the
cyclically time-expanded network on the one hand and the static
traffic assignment model on the other hand.  Basically, the demands
for different commodities\footnote{According to mathematical network flows, the term commodity denotes an origin-destination pair.} have to be subdivided to the number of
layers/time steps in the cyclically time-expanded network.

The cyclically time-expanded model
is a fully linear model with constant travel times on the expanded links. Also with the additional extensions in the following sections, the model is still a linear one. Thus, we can use standard
network flow algorithms to efficiently compute a system optimal traffic assignment, i.e., fast and exact
combinatorial and mathematical programming tools can be applied. A very fine time discretization of
time steps less than one second is easily achievable and quickly solvable even for large scenarios.

\begin{theorem}[\cite{KS10}]
 Using the cyclically time-expanded network the traffic assignment problem for a fixed traffic
signal coordination and for a fixed time granularity can be solved efficiently.
\end{theorem}

\begin{proof}
 The assignment problem can be formulated as a linear program (see~\cite{ahuja}),
because all travel times in the expanded network are constant. Since the LP has polynomial size with
respect to the input, i.e., the cyclically time-expanded network, it can be solved in polynomial time, e.g., by
using the ellipsoid method.
\end{proof}

\subsection{Detailed modeling of intersections} 

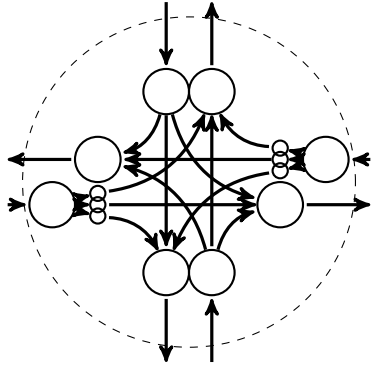
\begin{figure}[ht]
%\centering
\begin{center}
\begin{tikzpicture}
[scale=0.6,
vertex/.style={circle,draw,thick,inner sep=0pt,minimum size=2mm},
inv/.style={circle,draw,thick,inner sep=0pt,minimum size=0mm},
source/.style={circle,draw,thick,inner sep=0pt,minimum size=6mm},
sink/.style={rectangle,draw,thick,inner sep=0pt,minimum size=6mm},
vor/.style={-stealth',shorten >=1pt,very thick},
rueck/.style={stealth'-,shorten >=1pt,very thick},
demand/.style={},
unused/.style={-}
]

\node [source]	(v1) 	at (-3,-0.5) {};
\node [vertex]	(v1a) 	at (-2,-0.75) {}
 edge [rueck] node  {}	(v1);
\node [vertex]	(v1b) 	at (-2,-0.5) {}
 edge [rueck] node  {}	(v1);
\node [vertex]	(v1c) 	at (-2,-0.25) {}
 edge [rueck] node  {}	(v1);
\node [source]	(v2) 	at (-2,0.5) {};

\node [source]	(v3) 	at (3,0.5) {};
\node [vertex]	(v3a) 	at (2,0.75) {}
 edge [rueck] node  {}	(v3);
\node [vertex]	(v3b) 	at (2,0.5) {}
 edge [rueck] node  {}	(v3)
 edge [vor] node  {}	(v2);
\node [vertex]	(v3c) 	at (2,0.25) {}
 edge [rueck] node  {}	(v3);
\node [source]	(v4) 	at (2,-0.5) {}
 edge [rueck] node  {}	(v1b);

\node [source]	(v5) 	at (0.5,-2) {}
 edge [vor,bend right] node  {}	(v2)
 edge [vor,bend left] node  {}	(v4);
\node [source]	(v6) 	at (-0.5,-2) {}
 edge [rueck,bend right] node  {}	(v1a)
 edge [rueck,bend left] node  {}	(v3c);
\node [source]	(v7) 	at (-0.5,2) {}
 edge [vor,bend left] node  {}	(v2)
 edge [vor,bend right] node  {}	(v4)
 edge [vor] node  {}	(v6);
\node [source]	(v8) 	at (0.5,2) {}
 edge [rueck,bend left] node  {}	(v1c)
 edge [rueck,bend right] node  {}	(v3a)
 edge [rueck] node  {}	(v5);

\node [inv]	(iv1) 	at (-4,-0.5)  {}
  edge [vor] node  {}	(v1);
\node [inv]	(iv2) 	at (-4,0.5)  {}
  edge [rueck] node  {}	(v2);

\node [inv]	(iv3) 	at (4,0.5)  {}
  edge [vor] node  {}	(v3);
\node [inv]	(iv4) 	at (4,-0.5)  {}
  edge [rueck] node  {}	(v4);

\node [inv]	(iv5) 	at (0.5,-4)  {}
  edge [vor] node  {}	(v5);
\node [inv]	(iv6) 	at (-0.5,-4)  {}
  edge [rueck] node  {}	(v6);

\node [inv]	(iv7) 	at (-0.5,4)  {}
  edge [vor] node  {}	(v7);
\node [inv]	(iv8) 	at (0.5,4)  {}
  edge [rueck] node  {}	(v8);

\draw[dashed] (0,0) circle (3.65);
\end{tikzpicture}
\end{center}
  \caption[Expanded intersection]{Standard approach of an expanded intersection with arcs for each turning alternative. The incoming nodes of the horizontal road are subdivided into three nodes to model different lanes and queues for the turning directions. The vertical road is smaller, thus, all cars are waiting in the same queue.}
  \label{fig:crossex}
\end{figure}

\noindent
In Section~\ref{sec:cycmodel}, we modeled traffic signals by setting capacities of outgoing links to zero. At an intersection, we have to consider several lights and conflicts between crossing directions. %Instead of this
%fixed constraint, we can produce switchable traffic lights by multiplying the capacity with a binary
%decision variable. 
Thus, to model intersections with different lanes, turning directions, and
interior traffic signal offsets, we initially use a standard approach from
traffic networks. Before the time expansion is employed, every intersection node is split up into several
nodes for \emph{incoming} and \emph{outgoing traffic}; for each turning direction an interior arc connects one incoming node with one outgoing node (see Figure~\ref{fig:crossex}).  Each of these interior arcs is assigned
to one traffic light.

We will refer to the subset of all interior arcs of intersections by $E\subset A$. By choosing appropriate capacities $u(e)$ and transit times $t(e),~e\in E$, these interior links  
limit the outgoing flow of the waiting arcs of the several turning directions. Hence, we can also influence the creation of queues and back spill with these parameters.

\subsection{Modeling traffic signals}\label{sec:signals}

\noindent
Instead of setting fixed capacities to interior arcs of intersections, traffic signals can be modeled by binary decision
variables that switch the capacities of the interior arcs on or off by multiplying the capacity with an appropriate binary variable.  For each turning direction $e\in E$ we introduce $k$ binary variables
$b^e=(b^e_0,\dots,b^e_{k-1}) \in \{0,1\}^k$, one for each time step.  Now, we multiply the binary variables with the capacity constraint of the transit arcs corresponding to $e$ in the cyclically time-expanded network.  This yields $k$ inequalities for the flow in the expanded network, namely $f(e_t)\le b^e_t u(e_t) \quad \forall t \in \{0,\dots,k-1\}$. If $b^e_t=0$, then the effective capacity of $e_t$ is zero. Thus, turning direction $e$ cannot be used in time step $t$, i.e., the corresponding signal group shows a \emph{red} light. If $b^e_t=1$, then the capacity remains unchanged, hence we may think of a \emph{green} traffic light at time step $t$.

With these binary variables, we can permit or block flow on the interior arcs of an intersection. However, without further constraints, each signal group would be \emph{green} all the time. To ensure the interior logic of traffic lights at a single intersection, we have to couple these binary variables as in logic-based signal controls~\cite{Friedrich02}.  We concentrate on the following constraints:

\begin{itemize}
\item Each signal group should switch to \emph{green} only once per cycle.
\item Crossing directions must not have \emph{green} at the same time.
\item Some directions must have \emph{green} at the same time.
\item There has to be enough time to clear the intersection at phase changes before any other direction gets \emph{green}.
\item There has to be a minimum duration of each \emph{green} and \emph{red} phase.
\end{itemize}

In reality, many more constraints can occur, e.g., due to pedestrians. For simplicity, we only discuss here a small standard intersection with four legs and four signal groups $W,X,Y,Z$ as depicted in Figure~\ref{fig:kreuzung}. The common cycle time is $\Gamma$ and we use a cyclical expansion with $k$ time steps. Thus, in the following, all indices apply $\mathrm{modulo}~k$. As described above, we introduce $k$ binary variables $b^W = (b^W_0,\dots,b^W_{k-1})$ for signal group $W$ and similarly for $X,Y$, and $Z$ that represent the status of the corresponding light directly. We refer to these variables as \emph{status variables}.  

{\small
\begin{figure}[ht]
 \begin{center}
%    \psfrag{LSAA}{$W$}
%    \psfrag{LSAB}{$X$}
%    \psfrag{LSAC}{$Y$}
%    \psfrag{LSAD}{$Z$}
%    \psfrag{A}{$W$}
%    \psfrag{B}{$X$}
%    \psfrag{C}{$Y$}
%    \psfrag{D}{$Z$}
%    \psfrag{psi}{$\Psi_{Y/Z}$}  
%    \psfrag{cycle time}{cycle time $\Gamma$}%
   \includegraphics[width=4cm]{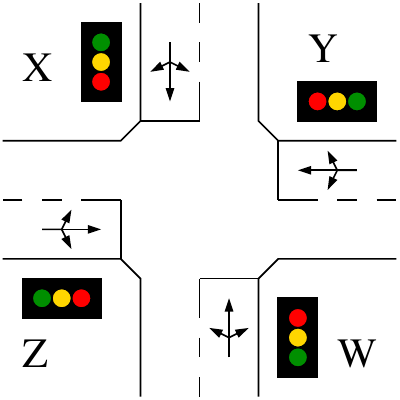}
   
   \caption[Simple intersection]{Simple intersection with four signal groups $W,X,Y$ and $Z$.}
   \label{fig:kreuzung}
 \end{center}
\end{figure}
}

Now, we demand that the signal group $W$ switches from \emph{red} to \emph{green} and from \emph{green} to \emph{red} only once during one cycle. To force this, we use $2k$ binary variables $b^{W,\textrm{on}} = (b^{W,\textrm{on}}_0,\dots,b^{W,\textrm{on}}_{k-1})$ and $b^{W,\textrm{off}} = (b^{W,\textrm{off}}_0,\dots,b^{W,\textrm{off}}_{k-1})$, respectively. These variables act like the decision variables for the offset in our previous model~\cite{KS10} and consequently, we refer to them as \emph{decision variables}. Thus, $b^{W,\textrm{on}}_i=1$ means that signal group $W$ switches from \emph{red} to \emph{green} at time step $i$. The following constraints fix the number of switching operations per cycle to 1.

\[ \sum_{i=1}^k b^{W,\textrm{on}}_i=1 \mbox{ and } \sum_{i=1}^k b^{W,\textrm{off}}_i=1 \]

Now, we link the decision variables to the status variables (remember to regard time $\textrm{modulo}~k$).

\[ b^W_i - b^W_{i-1} \le b^{W,\textrm{on}}_i~\forall i \in \{0,\dots,k-1\} \]

Thus, if $b^{W,\textrm{on}}_i=1$, then the signal may switch to \emph{green}, i.e., $b^W_{i-1}=0$ and $b^W_{i}=1$. If $b^{W,\textrm{on}}_{i}=0$, then $b^W_{i-1} \ge b^W_{i}$, that is, the status is not changed or the signal switches to red. Similarly, we formulate the constraints for switching the signal to \emph{red}.

\[ b^W_{i} - b^W_{i-1} \ge -b^{W,\textrm{off}}_{i}~\forall i \in \{0,\dots,k-1\} \]

However, it is not guaranteed that the signal group switches at all. We combine this with a constraint for minimum \emph{green} and \emph{red} periods. If the signal group should be \emph{green} for at least $g$ time steps, we require:

\[ \sum_{i=1}^{k} b^W_i \ge g. \]

If the signal group should show \emph{red} for at least $r$ time steps, we require:

\[ \sum_{i=1}^{k} b^W_i \le k-r. \] 

Therefore, if the signal group is \emph{green} for at least one time step and it is \emph{red} for at least one time step, two switching operations are forced. 

So far, we can switch traffic lights and we can control capacities. We continue with the safety constraints. Crossing directions must not have \emph{green} at the same time. However, \emph{red} for both directions is possible. Obviously, the following constraints ensure safe crossing for two signal groups $W$ and $X$ (cf.~Figure~\ref{fig:kreuzung}).

\[ b^W_i+b^X_i \le 1~\forall i \in \{0,\dots,k-1\} \]

Furthermore, we have to guarantee a clearance time. Together with the previous inequality, a clearance time of at least one time step is forced by the following constraint. 

\[ b^{W,\textrm{off}}_i + b^{X,\textrm{on}}_{i}+b^{X,\textrm{on}}_{i+1} \le 1 ~\forall i \in \{0,\dots,k-1\}\]

Thus, if signal group $W$ switches to \emph{red} in time step $i$, then $X$ cannot switch to \emph{green} at the same time step or at time step $i+1$. Longer clearance times are possible. One may even insert another phase. Hence, clearance time constraints have to be formulated for each pair of signal groups.  In contrast, we may also fix the clearance time and the phase order using the next equation.

\[b^{W,\textrm{off}}_{i} = b^{X,\textrm{on}}_{i+2}~\forall i \in \{0,\dots,k-1\}\]

Finally, if two signal groups should switch to \emph{green} (or \emph{red}) at the same time, e.g., group $W$ and $Y$ in Figure~\ref{fig:kreuzung}, we may use the following constraints.

\[b^{W,\textrm{on}}_{i}=b^{Y,\textrm{on}}_{i}~\forall i \in \{0,\dots,k-1\} \]

\subsection{Simultaneous optimization of assignment and signals}

\noindent
Each change in the traffic signal settings changes the travel times in the network. Thus, traffic signals and assignment have to be optimized simultaneously. We will refer to this problem as \textsc{Combined Traffic Assignment Traffic Signal Coordination} problem, i.e., find a traffic signal setting and an traffic assignment such that the overall travel time of all road users is minimized.

The choice of the objective is motivated by the following consideration. When optimizing the traffic assignment problem and
the traffic signal coordination simultaneously, measuring delays and stoppages does not characterize efficient coordinations.  
Traffic participants may choose arbitrary routes.  Thus, taking only
stops and delay into account, a road user may choose a very long detour
through the network just to avoid stopping or waiting in front of a
\emph{red} traffic light. This is rather unrealistic as most road users are
interested in the fastest way to their destination. Therefore, we use the total or average travel time (pure transit time + delay) as the measure of quality of the solution.

Summarizing the results of the previous subsection, we have introduced several linear equations or inequalities for the most important safety constraints at a traffic signal. %Their mode of action is mostly obvious. 
Let $G^T=(V^T,A^T,u^T)$ be a cyclically time-expanded traffic network and
commodities $\theta \in \Theta$, $\theta =
(s_\theta,z_\theta,d_\theta)$ (with origin $s_\theta\in V$, destination $z_\theta\in V$ , and demand $d_\theta\in \N$), capacities $u:A \rightarrow \N$, a set
$E \subset A$ of interior arcs at intersections with associated status variables $b^e$
for each $e \in E$, constant travel times $t(e)$ for each link $e\in A$ and flow functions
$f_\theta:A \rightarrow \R$ for each commodity. %Now, we can formulate a
%mixed integer program for the simultaneous optimization of traffic assignment and traffic signals.  

We extend a common multicommodity
min-cost circulation program (i.e., use a backward arc from $z_\theta$ to $s_\theta$ for each commodity $\theta \in \Theta$) for the cyclically time-expanded network
by adding the binary variables and capacity constraints above. Adding these binary variables and constraints to the linear program leads to the
following mixed integer program for the \textsc{Combined Traffic Assignment Traffic Signal Coordination} problem:

{\small \minProblem{
  \sum_{e \in A} \sum_{\theta \in \Theta} t_e f_{\theta}(e)\notag}{%
  &0 \le \sum_{\theta \in \Theta}  f_{\theta}(e)=f(e) \le u(e) & \forall ~ e \in
A \setminus E\label{eq:1}\\
  &\sum_{e \in \delta^+(v)} f_{\theta}(e) = \sum_{e \in \delta^-(v)}
f_{\theta}(e) & \forall ~ \theta \in \Theta \quad \forall ~ v \in
V\label{eq:2}\\
  &f_\theta((z_\theta,s_\theta)) = d_\theta  & \forall ~ \theta \in
\Theta\label{eq:3}\\
  %&\sum_{i=1}^{k} b^n_i = 1 &\forall ~ n \in \{1,\dots,N\}\label{eq:4}\\
  &f(e) \le b^e u(e) & \forall ~ e \in E \label{eq:5}\\
  &\textrm{Constraints of Section~\ref{sec:signals}} \label{eq:6} \\
  &f(e) \ge 0, ~ b^e \in \{0,1\}^k\notag}
}

The constraints of type~\eqref{eq:1} fix the capacity bounds,
type~\eqref{eq:2} implements the flow conservation and the
constraints~\eqref{eq:3} force the circulation. Equation~\eqref{eq:5} permits flow only on arcs that are switched on by the binary variables and in~\eqref{eq:6}, we insert all necessary constraints for these status variables. Due to the numerous equations for every intersection and every lane we omit the details here.

\subsection{Solving the MIP}

\noindent
The mixed integer linear program for the \textsc{Combined Traffic Assignment Traffic Signal Coordination} problem can be passed to every appropriate solver, for example CPLEX or GUROBI. Due to
the time-expansion and the decision variables, the MIP-formulation of
our model is rather large. In practice (see also Section~\ref{sec:results}), good solutions are found quite fast, but it takes a lot of time to close the gap between primal and dual solution. In some larger scenarios, a gap of 10 to 30 \% remains even after hours of computation. 

On the one hand, some further efforts are necessary to get fast and provable good solutions with help of the presented model. On the other hand, the \textsc{Combined Traffic Assignment Traffic Signal Coordination} problem is an $\cal NP$-hard problem and we may not hope for efficient algorithms (see, e.g,~\cite{GareyJohnson} for a guide to $\cal NP$-completeness). Even approximate solutions for a constant approximation ratio will be hard to get as shown in the following theorem.

\begin{theorem}\label{thm:complexity}
 For traffic signals which may only switch at discrete points in time, there exists no polynomial-time $\alpha$-approximation algorithm for any constant $\alpha\ge 1$ for the \textsc{Combined Traffic Assignment Traffic Signal Coordination} problem unless $\P=\NP$. 
\end{theorem}

\begin{proof}

We use a reduction from the $k$-\textsc{Directed Vertex Disjoint Paths} problem. This problem is defined as follows. Given a directed graph $G$ with $k$ terminal pairs $(s_1,t_1)$ to $(s_k,t_k)$, are there $k$ node-disjoint paths that connect the corresponding terminals? Even for $k = 2$, this problem is strongly $\NP$-complete~\cite{FHW80}.

Given an arbitrary instance of the $2$-\textsc{Directed Vertex Disjoint Paths}-problem, we construct the following combined traffic assignment traffic signal coordination problem. The network is formed by assigning unit capacities to the arcs, i.e., $u \equiv 1$, and zero transit times, i.e., $t_e=0 \quad \forall e\in E$. We use two commodities, each with demand $d_{1/2}=1$. Commodity $i$ ($i\in\{1,2\}$) starts at $s_i$ and ends at $t_i$. Each node of the graph is transformed to a signalized intersection and we assume that each traffic signal may only switch at integral points in time. At each signal, each turning direction has a separate traffic light and only one traffic light is allowed to be \emph{green} at each time step. That is, for all interior intersection arcs $\sum_{X} b^X_i\le1$ for all time steps $i$. Further, after each \emph{green} period of the signal there is a clearance time of at least $\lceil2\alpha\rceil$ time steps with all lights \emph{red}.  

Consequently, each single commodity can be routed in one time step. Due to the signal settings, commodities cannot split without waiting. Hence, both commodities can be routed in one time step if and only if there exist two vertex disjoint paths in the original network.  In contrast, if there exists no pair of vertex disjoint paths, both commodities have to meet at an intersection. They have to cross this intersection one after the other. Thus, half of the flow needs at least $\lceil2\alpha\rceil$ time steps to reach its destination. 
 
Therefore, any solution of the \textsc{Combined Traffic Assignment Traffic Signal Coordination} problem that needs less than $\lceil2\alpha\rceil$ time steps (or less than $\lceil\alpha\rceil$ total travel time) implies the existence of two vertex disjoint paths. This yields the claim.
\end{proof}

Obviously, the construction in the proof is somewhat extreme. Thus, one should not be discouraged by the above result, since one may get quite good solutions for realistic traffic networks in an acceptable time. However, even for traffic signals which may switch continuously, i.e., at arbitrary points in time, and without clearance times as well as planar networks with a maximum node degree of 4, the problem remains $\NP$-hard (see~\cite{meinediss}).  
Consequently, mixed integer programming is a reasonable approach for solving the \textsc{Combined Traffic Assignment Traffic Signal Coordination} problem, since we may not hope for an efficient algorithm unless $\P=\NP$. %We developed and applied several branching strategies and improved bounds to accelerate the solving process, but a detailed description of these approaches would  go beyond the scope of this paper.

\subsection{System optimum vs. user equilibrium}

\noindent
Since the famous example of Braess~\cite{braess}, one should be aware that road users may behave in a selfish way that is not optimal for the system as a whole.

However, up to here,  we only computed system optimal assignments based on the cyclically time-expanded model. This has two reasons. Firstly, we use a network with constant travel times. Thus, the system optimum is also a user equilibrium by definition. None of the road users can switch to a faster path towards the destination. Secondly, we use a network with capacities. Hence, \cite{CorreaSchulzStierMoses2004} implies that the user equilibrium is not unique anymore and that different equilibria may occur.

The theoretical results on the gap between system optimum and user equilibrium in static models are based upon properties of the link performance functions~\cite{roughgarden}. In the cyclically time-expanded model, these assumptions are not fulfilled. Instead, we have seen in Section~\ref{sec:performance} that we can capture much more details of the time dependent behavior. Thus, it is not possible to compare the (system optimal or user equilibrium) solutions of the static model and the cyclically time-expanded model directly.

Furthermore, we have seen non-convex and non-monotone travel time functions in Section~\ref{sec:performance}, which were created due to signals and platoons. Hence, considering signals and platoons in the static case with help of more sophisticated link performance functions would not meet the conditions which ensure the existence and uniqueness of an user equilibrium, neither.

Since we cannot provide a positive theoretical result here, we move the discussion of user equilibria to Section~\ref{sec:results} where we compare our system optimal solutions to equilibria obtained with a multi-agent simulation.

\section{Evaluation of real world scenarios}\label{sec:results}

% trc_signal_results.tex

\noindent In this section, we will use two real-world scenarios to demonstrate the capabilities of the cyclically time-expanded model. 
The first instance is a small part of the inner-city of Brunswick, Germany. Here, the focus is the optimization of offsets, split times and phase orders.
In the second scenario, the inner-city of Cottbus, Germany, we investigate the interplay of offset optimization and traffic assignment.
Afterwards, simulation tools are applied to study the gap between system optimum and user equilibrium.

\subsection{Brunswick}

\noindent
The first scenario consists of five signalized intersections and two pedestrian crossings in the inner-city of Brunswick. This scenario was provided by S\'{a}ndor Fekete from Technical University of Brunswick. It consists of the Bohlweg, a main arterial street through the city center connecting university and main station between Fallersleber Straße and Schloßplatz.

The network is presented in Figure~\ref{fig:brunswick_network}. We study three commodities visualized by colored arcs. The orange commodity and purple commodity have each twice the demand of the blue commodity. Since there are no alternative routes, route choice is of no importance in this scenario. As an additional constraint, the pedestrian signals~6 and~7 have to switch twice during one cycle of 84~seconds. Due to a nearby shopping mall and a tram station, a rather long minimum green time was assigned to these pedestrian signals. Consequently, these two signals will most likely be the bottleneck in this scenario.

\input{cyclic_results_bs1.tex}

\begin{figure}[ht!]
 \centering

 \begin{tikzpicture}[scale=0.02,
vertex/.style={circle,draw,thick,inner sep=0pt,minimum size=4mm,color=black,fill=white},
source/.style={circle,draw,thick,inner sep=0pt,minimum size=6mm},
sink/.style={rectangle,draw,thick,inner sep=0pt,minimum size=6mm,color=red,fill=white},
vor/.style={-stealth',shorten >=1pt,semithick},
comm1/.style={-stealth',shorten >=1pt,line width=3pt,color=orange},
comm2/.style={-stealth',shorten >=1pt,line width=3pt,color=blue},
comm3/.style={-stealth',shorten >=1pt,line width=3pt,color=purple}
]
\small
\node at (-20,0) {};

%axis
\begin{scope}[xshift=-15cm]
 \draw (0,0) -- (0,84);
 \draw[gray,opacity=0.5] (0,0) -- (480,0) -- (480,84) -- (0,84);
 \foreach \x in {6,12,18,24,30,36,42,48,54,60,66,72,78}
  \draw[gray,opacity=0.25] (0,\x) -- (480,\x);

 \draw (-2,0) -- (2,0);
 \draw (-2,84) -- (2,84);
 
 \node[left] at (-2,0) {0};
 \node[left] at (-2,84) {84};
\end{scope}

\begin{scope}[xshift=-15cm,yshift=110cm]
 \draw (0,0) -- (0,84);
 \draw[gray,opacity=0.5] (0,0) -- (480,0) -- (480,84) -- (0,84);
 \foreach \x in {6,12,18,24,30,36,42,48,54,60,66,72,78}
  \draw[gray,opacity=0.25] (0,\x) -- (480,\x);

 \draw (-2,0) -- (2,0);
 \draw (-2,84) -- (2,84);
 
 \node[left] at (-2,0) {0};
 \node[left] at (-2,84) {84};
\end{scope}

\begin{scope}[xshift=-15cm,yshift=220cm]
 \draw (0,0) -- (0,84);
 \draw[gray,opacity=0.5] (0,0) -- (480,0) -- (480,84) -- (0,84);
 \foreach \x in {6,12,18,24,30,36,42,48,54,60,66,72,78}
  \draw[gray,opacity=0.25] (0,\x) -- (480,\x);

 \draw (-2,0) -- (2,0);
 \draw (-2,84) -- (2,84);
 
 \node[left] at (-2,0) {0};
 \node[left] at (-2,84) {84};
\end{scope}

%Ampel1
\begin{scope}[xshift=0,yshift=0]
 \draw[color=green,fill=green] (0,0) rectangle (5,24);
 \draw[color=red,fill=red] (0,24) rectangle (5,84);

 \coordinate (v1a3) at (5,0);
 \coordinate (v1a4) at (5,24);
 
 \node at (2,-10) {1};
 
\end{scope}

%Ampel2
\begin{scope}[xshift=80cm,yshift=0]
 \draw[color=green,fill=green] (0,8) rectangle (5,74);
 \draw[color=red,fill=red] (0,74) rectangle (5,84);
 \draw[color=red,fill=red] (0,0) rectangle (5,8);

 \coordinate (v2e3) at (0,8);
 \coordinate (v2e4) at (0,32);
 
 \coordinate (v2a3) at (5,8);
 \coordinate (v2a4) at (5,32);
 
 \node at (2,-10) {2};

\end{scope}

%Ampel3a
\begin{scope}[xshift=230cm,yshift=0]
 \draw[color=red,fill=red] (0,0) rectangle (5,23);
 \draw[color=green,fill=green] (0,23) rectangle (5,47);
 \draw[color=red,fill=red] (0,47) rectangle (5,84);
 
 \coordinate (v3e1) at (0,23);
 \coordinate (v3e2) at (0,47);

 \coordinate (v3a1) at (5,23);
 \coordinate (v3a2) at (5,47);
 
 \node at (3,-10) {3A};
  
\end{scope}

%Ampel6
\begin{scope}[xshift=330cm,yshift=0]
 \draw[color=red,fill=red] (0,13) rectangle (5,33);
 \draw[color=green,fill=green] (0,33) rectangle (5,63);
 \draw[color=red,fill=red] (0,63) rectangle (5,84);
 \draw[color=green,fill=green] (0,0) rectangle (5,12);
 \draw[color=red,fill=red] (0,12) rectangle (5,13);
 
 \coordinate (v4e1) at (0,33);
 \coordinate (v4e2) at (0,57);

 \coordinate (v4a1) at (5,33);
 \coordinate (v4a2) at (5,57);

 \node at (2,-10) {6};
 
\end{scope}

%Ampel7
\begin{scope}[xshift=380cm,yshift=0]
 \draw[color=red,fill=red] (0,17) rectangle (5,38);
 \draw[color=green,fill=green] (0,38) rectangle (5,68);
 \draw[color=red,fill=red] (0,68) rectangle (5,84);
 \draw[color=red,fill=red] (0,0) rectangle (5,5);
 \draw[color=green,fill=green] (0,5) rectangle (5,17);

  \coordinate (v5e1) at (0,38);
 \coordinate (v5e2) at (0,62);

 \coordinate (v5a1) at (5,38);
 \coordinate (v5a2) at (5,62);

 \node at (2,-10) {7};

\end{scope}

%Ampel4
\begin{scope}[xshift=440cm,yshift=0]
 \draw[color=green,fill=green] (0,0) rectangle (5,23);
 \draw[color=red,fill=red] (0,23) rectangle (5,44);
 \draw[color=green,fill=green] (0,44) rectangle (5,84);
 
  \coordinate (v6e1) at (0,44);
 \coordinate (v6e2) at (0,68);

 \node at (2,-10) {4A};
  
\end{scope}

%Ampel3b
\begin{scope}[xshift=230cm,yshift=110cm]
 \draw[color=green,fill=green] (0,0) rectangle (5,21);
 \draw[color=red,fill=red] (0,21) rectangle (5,49);
 \draw[color=green,fill=green] (0,49) rectangle (5,84);
 
  \coordinate (v3e3) at (0,47);
  \coordinate (v3e4) at (0,84);
  \coordinate (v3e5) at (0,0);
  \coordinate (v3e6) at (0,2);

 \coordinate (v3a3) at (5,49);
 \coordinate (v3a4) at (5,74);
 \coordinate (v3a5) at (5,84);
 
 \coordinate (v3a6) at (5,0);
 \coordinate (v3a7) at (5,2);

 \node at (2,-10) {3B};

\end{scope}

%Ampel6 lvl2
\begin{scope}[xshift=330cm,yshift=110cm]
 \draw[color=red,fill=red] (0,13) rectangle (5,33);
 \draw[color=green,fill=green] (0,33) rectangle (5,63);
 \draw[color=red,fill=red] (0,63) rectangle (5,84);
 \draw[color=green,fill=green] (0,0) rectangle (5,12);
 \draw[color=red,fill=red] (0,12) rectangle (5,13);

 \coordinate (v4e3) at (0,57);
 \coordinate (v4e4) at (0,84);
 \coordinate (v4e5) at (0,0);
 \coordinate (v4e6) at (0,10);
 \coordinate (v4e7) at (0,12);

 \coordinate (v4a3) at (5,59);
 \coordinate (v4a4) at (5,63);
 
 \coordinate (v4a5) at (5,0);
 
 \coordinate (v4a8) at (5,12);
 
 \node at (2,-10) {6};
 
\end{scope}

%Ampel7 lvl2
\begin{scope}[xshift=380cm,yshift=110cm]
 \draw[color=red,fill=red] (0,17) rectangle (5,38);
 \draw[color=green,fill=green] (0,38) rectangle (5,68);
 \draw[color=red,fill=red] (0,68) rectangle (5,84);
 \draw[color=red,fill=red] (0,0) rectangle (5,5);
 \draw[color=green,fill=green] (0,5) rectangle (5,17);

   \coordinate (v5e3) at (0,64);
 \coordinate (v5e4) at (0,68);
 
   \coordinate (v5e5) at (0,5);
  \coordinate (v5e8) at (0,17);

  \coordinate (v5a3) at (5,64);
 \coordinate (v5a4) at (5,68);
 
    \coordinate (v5a5) at (5,5);
 \coordinate (v5a10) at (5,17);
 
 \node at (2,-10) {7};

\end{scope}

%Ampel4 lvl2
\begin{scope}[xshift=440cm,yshift=110cm]
 \draw[color=green,fill=green] (0,0) rectangle (5,23);
 \draw[color=red,fill=red] (0,23) rectangle (5,44);
 \draw[color=green,fill=green] (0,44) rectangle (5,84);
 
   \coordinate (v6e3) at (0,70);
 \coordinate (v6e4) at (0,74);
 
     \coordinate (v6e5) at (0,11);
  \coordinate (v6e10) at (0,23);
 
 \node at (2,-10) {4A};
  
\end{scope}

%Ampel4b
\begin{scope}[xshift=440cm,yshift=220cm]
 
 \draw[color=green,fill=green] (0,0) rectangle (5,7);
 \draw[color=green,fill=green] (0,44) rectangle (5,84);
 \draw[color=red,fill=red] (0,7) rectangle (5,44);
  \coordinate (v6a1) at (0,44);
  \coordinate (v6a2) at (0,49);
 \coordinate (v6a3) at (0,84);
 \coordinate (v6a4) at (0,0);
 \coordinate (v6a5) at (0,7);
 
 \node at (2,-10) {4B};
  
\end{scope}

%Ampel5a
\begin{scope}[xshift=120cm,yshift=110cm]
 \draw[color=red,fill=red] (0,0) rectangle (5,36);
 \draw[color=green,fill=green] (0,36) rectangle (5,75);
 \draw[color=red,fill=red] (0,75) rectangle (5,84);
 
 \coordinate (v8a1) at (5,36);
 \coordinate (v8a2) at (5,73);
 \coordinate (v8a3) at (5,75);

\node at (2,-10) {5A};
  
\end{scope}

%Ampel5b
\begin{scope}[xshift=120cm,yshift=220cm]
 \draw[color=green,fill=green] (0,0) rectangle (5,33);
 \draw[color=red,fill=red] (0,33) rectangle (5,78);
 \draw[color=green,fill=green] (0,78) rectangle (5,84);
 
  \coordinate (v8e1) at (5,79);
  \coordinate (v8e2) at (5,84);
  \coordinate (v8e3) at (5,0);
  \coordinate (v8e4) at (5,34);
  \coordinate (v8e5) at (5,41);
  
\node at (2,-10) {5B};
\end{scope}

%comm1
%\fill[fill=orange,opacity=0.5] (v1a1) -- (v2e1) -- (v2e2) -- (v1a2);
\fill[fill=orange,opacity=0.5] (v1a3) -- (v2e3) -- (v2e4) -- (v1a4);

%\fill[fill=orange,opacity=0.5] (v2a1) -- (v3e1) -- (v3e2) -- (v2a2);
\fill[fill=orange,opacity=0.5] (v2a3) -- (v3e1) -- (v3e2) -- (v2a4);

\fill[fill=orange,opacity=0.5] (v3a1) -- (v4e1) -- (v4e2) -- (v3a2);

\fill[fill=orange,opacity=0.5] (v4a1) -- (v5e1) -- (v5e2) -- (v4a2);

\fill[fill=orange,opacity=0.5] (v5a1) -- (v6e1) -- (v6e2) -- (v5a2);

%comm2

\fill[fill=blue,opacity=0.5] (v8a1) -- (v3e3) -- (v3e4) -- (v8a2);
\fill[fill=blue,opacity=0.5] (v8a2) -- (v3e5) -- (v3e6) -- (v8a3);

\fill[fill=blue,opacity=0.5] (v3a3) -- (v4e3) -- (v4e4) -- (v3a4);
\fill[fill=blue,opacity=0.5] (v3a4) -- (v4e5) -- (v4e6) -- (v3a5);
\fill[fill=blue,opacity=0.5] (v3a6) -- (v4e6) -- (v4e7) -- (v3a7);

\fill[fill=blue,opacity=0.5] (v4a3) -- (v5e3) -- (v5e4) -- (v4a4);

\fill[fill=blue,opacity=0.5] (v5a3) -- (v6e3) -- (v6e4) -- (v5a4);

\fill[fill=blue,opacity=0.5] (v4a5) -- (v5e5) -- (v5e8) -- (v4a8);

\fill[fill=blue,opacity=0.5] (v5a5) -- (v6e5) -- (v6e10) -- (v5a10);

%\fill[fill=blue,opacity=0.5] (v4a7) -- (v5e7) -- (v5e8) -- (v4a8);

%\fill[fill=blue,opacity=0.5] (v5a7) -- (v6e7) -- (v6e8) -- (v5a8);

%\fill[fill=blue,opacity=0.5] (v5a9) -- (v6e9) -- (v6e10) -- (v5a10);

%comm3
\fill[fill=purple,opacity=0.5] (v6a2) -- (v8e3) -- (v8e4) -- (v6a3);
\fill[fill=purple,opacity=0.5] (v6a1) -- (v8e1) -- (v8e2) -- (v6a2);
\fill[fill=purple,opacity=0.5] (v6a4) -- (v8e4) -- (v8e5) -- (v6a5);

\end{tikzpicture}
 \caption{Optimized green bands for the Brunswick scenario. Time and state of the signals is shown on the vertical axis. Signals are labelled with intersection numbers and signal groups, e.g., signal $3A$ and $3B$ must not be \emph{green} at the same time. Distances on the horizontal axis are chosen with respect to transit time, the slope of the parallelograms is chosen with respect to free speed. Note that the purple commodity is travelling from right to left and the cyclic overflow is visualized by a negative slope.}\label{fig:brunswick_solution}
 
\end{figure}
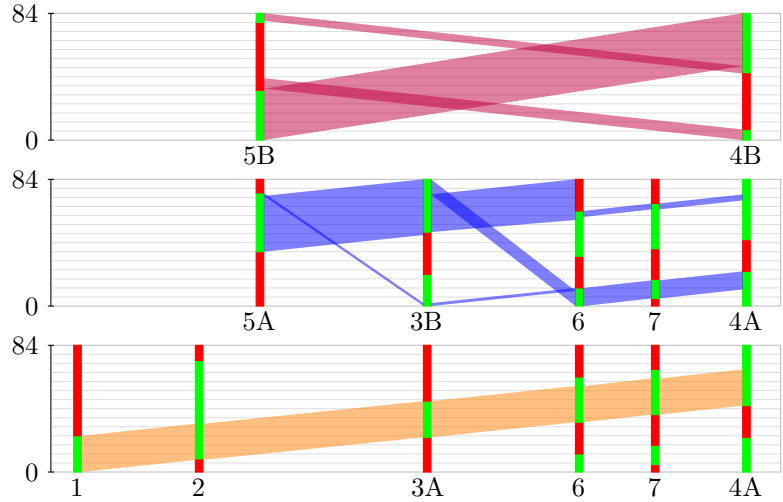

Usually, the optimal solution for instances of this scenario in our framework is found and proved within a few seconds. Solution time depends on demands, signal parameters, optimization parameters in CPLEX or Gurobi, respectively, and destroying symmetry of the cyclic network, e.g., by fixing an offset of one intersection. An exemplary solution is shown in Figure~\ref{fig:brunswick_solution}.

The  path-time diagrams in Figure~\ref{fig:brunswick_solution} indicate that road users on the Bohlweg (orange) may pass every signal with a speed of 30 miles per hour (50 km/h) without stopping. Also for the purple commodity, the second signal becomes green exactly at the arrival of the first flow. The main bottleneck are the pedestrian crossings. Here, the blue commodity has to wait at signal 6, but signals 3, 7, and 4 can be passed without stopping. Green splits are automatically adapted to flow values. Summarizing, our approach can be used to optimize traffic signal settings and to find good coordinations of signals.

\subsection{Cottbus}

\noindent
Whereas the first example does not allow any route choice, the second scenario is based on a much larger network. Cottbus is a German town just south of Berlin with about 100,000 inhabitants. The scenario consists of the whole inner-city with an area of 10 square kilometers and 32 signalized intersections. The scenario is presented in Figures~\ref{fig:umlegung_visum}~to~\ref{fig:wartezeiten}.

The network itself was created with data from \url{www.openstreetmap.org}. Very small streets, traffic-calmed areas, dead ends, etc. were removed. Traffic demand, i.e, origin-destination pairs, was generated based on anonymised data obtained from the German employment agency. We have access to aggregated home addresses and workplaces of all employees in the area of Cottbus and the surrounding rural district. This allows a good approximation of traffic in the morning and afternoon peak. We reduced the number of commodities by aggregating commodities with similar origins and destinations. That is, we merged all points of interest within a certain neighborhood to a single terminal. Furthermore, we only consider commodities where the shortest path connection passes through the city center. This yields about 250 commodities.

For our first experiments, we used 17  out of these 250 commodities to save computation time. However, demand was scaled back to 100 percent. We also limited our model to offset optimization, all split times and phase orders were fixed to measured values. We started with an already optimized signal setting for a fixed assignment, i.e.~the signals provide minimum delay for this fixed assignment. This assignment was computed by VISUM (ptv AG, \url{http://www.ptvgroup.com}) and we applied our mixed integer program without rerouting to optimize the signal settings. Afterwards, our approach could reduce waiting time in the network by another 11 \% by rerouting traffic flow and resetting traffic signals.

An interesting property of our optimized solutions is the creation of circulations around the traffic center (see Figure~\ref{fig:umlegung_uns}). This omits a lot of conflicts between opposing traffic streams and allows progressive signal settings for many road users. The computational results were verified by simulation with VISSIM (ptv AG, \url{www.ptvgroup.com}) and MATSim (\url{www.matsim.org}).

\input{cyclic_results_cb1.tex}

\input{cyclic_results_cb2.tex}

\input{cyclic_results_cb3.tex}

\input{cyclic_results_cb4.tex}

Unfortunately, a gap of 20 percent between the obtained solution and the lower bound remains when solving this scenario with CPLEX or Gurobi even after hours of computation. Hence, we cannot prove optimality and 20 percent leaves space for improvements. However, experiments with smaller scenarios showed that the dual bounds are much harder to compute. That is, the primal solution converges faster towards the optimal solution than the dual bound does. Thus, the gap between the obtained solution and the optimal solution is most likely much smaller than 20 percent.

In contrast, other approaches for traffic signal optimization and traffic assignment use genetic programming or similar heuristics. Thus, these approaches do not provide such duality gaps at all.

Concluding, one of the main intentions of our model -- feedback between traffic
assignment and coordination during the optimization process -- is achieved.  A conventional approach would first determine the
assignment, afterwards it would optimize the traffic signals. This could lead to a very uniform distribution of traffic in the network. 

In contrast, the computed solutions of the simultaneous optimization with our model are quite
different. Traffic is assigned to a few roads. Consequently, the flow rate is high on these roads. However, these routes automatically get very high priority in the signal optimization subproblem and traffic is concentrated in a small number of dense platoons.  In artificial scenarios~\cite{meinediss}, we could save up to 75 percent waiting time by simultaneous optimization compared to the consecutive approach.

These completely different but high-performance assignments
characterize the presented model.  We believe that similar efficient results cannot be obtained without simultaneously considering assignment and
coordination, i.e., a highly-adaptive traffic signal optimization will have
problems to find a competitive solution.  These results also suggest that
traffic signal coordination could be used to actively route or
redirect traffic.

\subsection{System optimum vs. user equilibrium revisited}

\noindent
In this section, we will compare the computational results of the cyclically time-expaned network model with simulation results obtained with MATSim. 

MATSim is a multi-agent traffic simulation. It can simulate user equilibrium traffic flows with an iterative approach consisting of simulation and replanning rounds. Therefore, each agent has its own set of plans. A single plan consists of activities, routes, and departure times. Furthermore, a plan executed in a simulation round is evaluated and gets an individual score. This score includes travel times and other parameters with arbitrary weighting. New plans can also be generated via shortest paths calculations in the network with the current travel times. In the replanning phase, the plans for execution in the next simulation round are chosen. A certain percentage of the agents simply chooses the plans with the best score so far, whereas the other agents try a randomly chosen alternative plan. After several rounds of simulation and replanning, this process converges to a stable state. Since each agent has its own set of plans with individual scores, this state can be accepted as a user equilibrium.

We consider the Cottbus network. Out of the 250 commodities, we chose 54 commodities with the highest demands which add up to about 50 \% of the total demand. At first, we optimized traffic and signals with our approach. Afterwards, we used the obtained signal settings and the same origin-destination pairs to compute user equilibrium assignments with the multi-agent simulation MATSim. Furthermore, we fixed traffic signal offsets to random values to compare assignments only. Hence, the base for the results presented in Table~\ref{tab:matsim} differs in two aspects.  Firstly, we compare calculated results to measured results. Secondly, we compare system optima to user equilibria.  

\begin{table}[ht!]\centering
%  \begin{tabbing}
%  Szenario~~~~~~~~~~ \= cyclically expanded network~~ \= MATSim simulation~~ \= Gap (\%)\=~\kill
%  scenario      \> cyclically expanded network\>MATSim simulation\> Gap (\%)\\
%  opt          \> \>\makebox[0cm][r]{1,103,380} \>  \makebox[0cm][r]{1.140.077}\> \makebox[0cm][r]{3.33}\\ 
%  best random  \> 1,156,888 \>1,173,303 \> 1.41\\ 
%  med random   \> 1,197,666 \>1,210,419 \> 1.06 \\
%  avg random   \> 1,198,688 \>1,236,851 \> 3.18\\ 
%  worst random \> 1,285,766 \>1,291,993 \> 1.00%\\ 
%  %base case:    \>  \>1.172.967 
% \end{tabbing}
 \begin{tabular}{lrrr}
 \toprule
 scenario   &   cyclically expanded network&MATSim simulation& Gap (\%)\\
 \midrule
 opt          & 1,103,380 &1,140,077 & 3.33\\ 
 best random  & 1,156,888 &1,173,303 & 1.41\\ 
 med random   & 1,197,666 &1,210,419 & 1.06 \\
 avg random   & 1,198,688 &1,236,851 & 3.18\\ 
 worst random & 1,285,766 &1,291,993 & 1.00\\ 
 %base case:    \>  \>1.172.967 
 \bottomrule
\end{tabular}
\caption{Comparison of system optimal solutions of the cyclically time-expanded network model and user equilibrium solutions of MATSim. All values represent total travel times in seconds for the morning peak from 6:30 to 9:30 am. Besides the optimized traffic signals (opt), we also compare optimized assignments for 100 random traffic signal settings.}\label{tab:matsim}
\end{table}

First of all, the optimized solution and the random signal settings appear in the same order for both calculation and simulation. But surprisingly, the difference between the travel times in the cyclically time-expanded network and in MATSim is really small, i.e., one to four percent. Since no calibration was applied and some simplifying assumptions were made in the developing of the cyclically time-expanded model, this suggests that the gap between system optimum and user equilibrium is almost negligible in this scenario. 

Furthermore, the optimized signal settings of the system optimal assignment also improve the value of the user equilibrium by three percent compared to the best random solution.  Thinking in absolute values, optimized offsets can already save about 10 hours of delay in every peak compared to a good random traffic signal setting in this scenario. Extrapolating this result to full demand, two peaks per day, and 250 working days per year, this adds up to about 10.000 hours per year. Even for a small city like Cottbus, where traffic congestion is not a serious problem, this is a worthwhile economic gain. 

Additionally, the consecutive approach of first computing an assignment and afterwards optimizing traffic signals which was used as our benchmark described in the previous subsection yields a total travel time of 1,172,967 seconds. Thus, the consecutive approach is worse than the best random guess.

% \section{Discussion}
% 
% 
% \input{trb_signal_discussion.tex}

\section{Conclusions}

% trc_signal_conclusion.tex
\noindent
In this paper, we focused on travel times in our cyclically time-expanded network. As a main insight, we have shown that even a completely linear model is capable of creating complex travel time functions. This is achieved at the cost  of a much higher dimension, i.e, much more variables are used to describe traffic flow on a single link. Simulation results show that the cyclically time-expanded model is realistic and can capture much more properties of inner-city traffic than static link performance functions can do. 

Even better, the cyclically time-expanded model can be used to optimize traffic assignment and traffic signal settings simultaneously. This yields very efficient solutions compared to iterative approaches. Furthermore, the model allows strict mathematical programming. Thus, we can use solvers like CPLEX, which provide a guarantee on the optimality of a solution. This is a considerable advantage compared to other approaches like TRANSYT and others, which use genetic programming or similar heuristics. These tools produce very good solutions, but they do not provide an estimate of the gap towards the optimal solution.

After this proof of concept, our further research focuses on the applicability of our model. Thus, we will try to develop  branch\&bound strategies or to adapt cutting plane techniques to accelerate the optimization process. There are some other shortcomings in the modeling. For example, we cannot guarantee the First In First Out property in every situation. Besides this rather technical problems, studying user equilibria in our model is also interesting from a theoretical as well as from a practical point of view. One main question is the influence of traffic signal optimization on user equilibria. In a network without signals, system optimum and user equilibrium may look quite different. Is our method of simultaneously optimizing signals and assignment capable of pushing the equilibrium flow towards the system optimum flow by privileging the routes of the system optimum with progressive signal settings and penalizing the bad links of the user equilibrium with high waiting times?

\section*{Acknowledgments}
The authors thank Gregor Wünsch and Christian Liebchen who were involved in the first ideas of the cyclically time-expanded model. We also thank Kai Nagel and his group at TU Berlin as well as Klaus Nökel (ptv AG) for providing various insights into traffic simulation concepts and the corresponding simulation tools. This work was supported by the German Research Foundation (DFG, grant number KO 2256/2-1).

\bibliographystyle{plain}
%\bibliography{trc_bib.bib}

\end{document}